\documentclass[a4paper,12pt]{article}
\usepackage[dvips]{graphics}
\usepackage{epsfig}
\usepackage{psfrag}
\usepackage[psamsfonts]{amssymb}
\usepackage{amsmath}
\usepackage{indentfirst}
\usepackage{amssymb}
\usepackage{wrapfig}

\usepackage{graphicx}
\usepackage{amsmath}
\usepackage{amssymb}
\usepackage{theorem}
\usepackage{url}

\renewcommand{\thetheorem}{\arabic{section}.\arabic{theorem}}

\renewcommand{\thetheorem}{\arabic{section}.\arabic{theorem}}
\newtheorem{definition}{Definition}
\renewcommand{\thetheorem}{\arabic{section}.\arabic{theorem}}

\renewcommand{\thetheorem}{\arabic{section}.\arabic{theorem}}

\renewcommand{\thetheorem}{\arabic{section}.\arabic{theorem}}

\renewcommand{\thetheorem}{\arabic{section}.\arabic{theorem}}
\newtheorem{assertion}{Assertion}
\renewcommand{\thetheorem}{\arabic{section}.\arabic{theorem}}

\begin{document}

\rule{0mm}{50mm}

\centerline{\Large A simple microstructure return model}
\smallskip
\centerline{\Large explaining microstructure noise and Epps effects}

\rule{0mm}{20mm}

\centerline{\large A. Saichev$^{1,3}$, D. Sornette$^{1,2}$}

\rule{0mm}{20mm}

\centerline{$^{1}$\footnotesize ETH Zurich -- Department of Management, Technology and Economics, Switzerland}

\centerline{$^{2}$\footnotesize Swiss Finance Institute, 40, Boulevard du Pont-d' Arve, Case Postale~3, 1211 Geneva 4, Switzerland}

\centerline{$^{3}$\footnotesize Nizhni Novgorod State University -- Department of Mathematics, Russia.}

\rule{0mm}{60mm}

\centerline{E-mail addresses: saichev@hotmail.com \& dsornette@ethz.ch}

\clearpage

\rule{0mm}{30mm}

\centerline{\Large A simple microstructure return model}
\smallskip
\centerline{\Large explaining microstructure noise and Epps effects}

\rule{0mm}{10mm}

\begin{abstract}
We present a simple microstructure model of financial returns that combines
(i) the well-known ARFIMA process applied to tick-by-tick returns, (ii) the
bid-ask bounce effect, (iii) the fat tail structure of the distribution of returns
and (iv) the non-Poissonian statistics of inter-trade intervals. This model allows us
to explain both qualitatively and quantitatively
important stylized facts observed in the statistics of microstructure returns, including
the short-ranged correlation of returns, the long-ranged correlations of absolute returns,
the microstructure noise and Epps effects. 
According to the microstructure noise effect, volatility 
is a decreasing function of the time scale used to estimate it. Paradoxically,
the Epps effect states that cross correlations between asset returns
are increasing functions of the time scale at which the returns are estimated.
The microstructure noise is explained as the 
result of the negative return correlations inherent in the
definition of the bid-ask bounce component (ii). In the presence of a genuine
correlation between the returns of two assets, 
the Epps effect is due to an average statistical
overlap of the momentum of the returns of the two
assets defined over a finite time scale in the presence of the long
memory process (i).   
\end{abstract}

\rule{0mm}{10mm}

Didier Sornette

Department of Management, Technology and Economics

(D-MTEC, KPL F38.2) ETH Zurich

Kreuzplatz 5

CH-8032 Zurich

Switzerland

\clearpage

\smallskip

\section{Introduction}

In the last decade, a lot of attention has been directed towards gathering empirical facts and
developing theoretical understanding of the microstructure (i.e., on tick-by-tick scales)
characterizing the behavior of stocks prices, their corresponding returns and volatility (see for instance \cite{Ait2009, Bandi2006,Bandi2008,Bouchaud2006, Cont2010, Munnix2010, Voev2007, Zhang2005}).
These investigations have revealed seemingly contradictory stylized facts
in the statistical behavior of the dynamics of prices and returns. Let us mention the
well-known short-ranged correlations of returns together with the long-ranged correlations of
the absolute value of returns \cite{Cont2000,Ding1993},
the microstructure noise effect of realized volatility \cite{Ait2009,Bouchaud2006,Zhang2005} and
the Epps effect  of cross-correlations of returns \cite{Bouchaud2006,Epps1979}.
The microstructure noise effect  refers to the observation that volatility 
is a decreasing function of the time scale used to estimate it. In contrast, the
Epps effect describes the fact that cross correlations between asset returns
are increasing functions of the time scale at which the returns are estimated.
These stylized facts are observed in a large variety of financial markets,
including stocks, futures and options and they are thus considered as universal
properties that are inherent to high-frequency financial markets data. However,
there is still no theoretical model that can account simultaneously for the four mentioned
stylized facts (see however Ref.\cite{Munnixetal2011} which emphasizes that the
asynchrony of trades as well as the decimalization
of stock prices are large contributors to the Epps effect. This paper also
contains a review of previous papers attempting to explain the Epps effect).
It is the purpose of the present paper to fill this gap and propose
 simple model for the microstructure of returns that describes these stylized facts
both at a qualitative and quantitative level. 

Our model of tick-by-tick returns contains the following ingredients:
(i) a long-memory centered Gaussian variable following an ARFIMA process;
(ii) a sign bounce generalizing the bid-ask bounce effect
governed by a quasi-periodic Bernouilli random variable;
(iii) a random amplitude drawn from a fat-tail distribution;
and (iv) the specification of the iid sequence of inter-trade intervals
distributed according to a given probability density function chosen
as a simple Weibull or a generalized gamma distribution.
We show that the microstructure noise
effect results simply from the negative return correlations inherent in the
definition of the sign bounce component (ii). In the presence of a genuine
correlation between the returns of two assets, 
the Epps effect appears due to an average statistical
overlap of the momentum of the returns of the two
assets defined over a finite time scale in the presence of the long
memory process (i). In our microstructure framework,
correlations between two assets correspond to buy or sell
pressure exerted on both of them, but not exactly at the same time
(except if the correlation is 1), due to possible delays in 
implementation as well as herding effects between trading decisions
resulting from human decisions or algorithmic high-frequency trading.
We stress that, notwithstanding this long memory (i), the sign bounce
process (ii) ensures that auto-correlations of returns are short-lived, in 
agreement with empirical evidence. 
 
The presentation is organized as follows. Section~2 describes in details our model
and derives the quantitative characteristics of the long ranged correlations of
the absolute tick-by-tick returns. Section~3 discusses returns correlations
in calendar time scale. Significant attention is given to the statistical description
of inter-trade time intervals. Section~4 describes quantitatively
the statistical properties of the microstructure noise effect in the frame of the proposed model.
Then, Section 5 reveals the roots of the Epps effect, as suggested from our microstructure model.
Two appendices provide detailed information on the derivation of the
analytic expressions used in the main text.

\section{Model for the microstructure of financial returns at the tick scale}

\subsection{Instantaneous returns vs tick-by-tick returns}

Let $L(t)$ be the log-price of some asset. The
corresponding return $R_\Delta(t)$ as time scale $\Delta$ is defined
as the increment of the log-price over the time interval $(t-\Delta,t)$
\begin{equation}
R_\Delta(t) = L(t) - L(t-\Delta) ~.
\end{equation}
In the present work, we develop a phenomenological theory of returns $R_\Delta(t)$
based on the \emph{instantaneous return} $R(t)$ defined as the derivative of log-price $L(t)$:
\begin{equation}
R(t) = \frac{d L(t)}{dt} ~.
\end{equation}
Knowing the instantaneous returns, the $\Delta$-scale returns $R_\Delta(t)$
are obtained in integral form
\begin{equation}\label{rtdeldef}
R_\Delta(t) = \int^t_{t-\Delta} R(t') dt' ~.
\end{equation}

We consider the situation where the instantaneous returns
are delta-pulses of the form
\begin{equation}\label{returnsdef}
R(t) = \sum_{k} r_k \delta(t-t_{k})~ .
\end{equation}
Expression (\ref{returnsdef}) reflects the discrete nature of the
process generating changes of log-prices, which occur by jumps $\{r_k\}$ at the
ordered tick-by-tick instants $\{t_k\}$:
\begin{equation}\label{tkordering}
\dots < t_{k-1} < t_k < t_{k+1} \dots \qquad (t_0 <0 , ~ t_1>0 ) ~.
\end{equation}

%The ticks-by-tick returns $\{r_k\}$ are referred to as ``log-price jumps''.

We suggest below that the sequences $\{t_k\}$ and $\{r_k\}$ are mutually statistically independent.
Moreover, we assume that the sequences $\{t_k\}$ and $\{r_k\}$ are stationary in the sense that
the statistical properties of inter-trade intervals
\begin{equation}\label{taukmdef}
\tau_{k,m} = t_{k+m} - t_k
\end{equation}
and of pairs $(r_{k+m},r_{k})$ depend only on $m$.

The instantaneous return $R(t)$ and  $\Delta$-scale returns $R_\Delta(t)$ are
defined in the \emph{calendar time scale}, where the time is measured in units of seconds, hours, years,
and so on. In particular, expressions \eqref{rtdeldef} and \eqref{returnsdef} are representations of returns and of instantaneous returns in the calendar time scale.
Sometimes, it may be more convenient to use the \emph{tick-by-tick time scale}, where one measures time in subsequent trade numbers. In this sense, the sequence $\{r_k\}$ is the representation of
instantaneous returns in the tick-by-tick time scale. Hereinafter, we shall discuss in details
the interrelation between the instantaneous return $R(t)$ given by \eqref{returnsdef}
defined in calendar time and the returns $\{r_k\}$ in tick-by-tick time.

\subsection{Formulation of the model of tick-by-tick returns}

Let us now define a simple stochastic model of tick-by-tick returns $\{r_k\}$ of the form
\begin{equation}\label{tbtrmod}
r_k : = X_k M_k / H_k~,
\end{equation}
which captures both the short-range correlation of returns and the
long-range correlation of the absolute value of returns.
The sequence $\{X_k\}$ is generated
by a standard ARFIMA process \cite{Granger1980,Hosking1981}
satisfying the discrete fractional difference equation
\begin{equation}\label{fracxeq}
(1-T)^d X_k =  \Upsilon \cdot u_k , \qquad d\in (0,1/2)~,
\end{equation}
where $T$ is the unit tick-lag operator, the $u_k$'s are iid random
variables with zero mean and unit variance,
$\Upsilon$ is a normalizing constant
such that $\rho_0=1$ where $\rho_m$ is the correlation function of the ARFIMA process
\begin{equation}\label{rhmdef}
\rho_m := \text{E}\left[X_k X_{k+m}\right]~.
\end{equation}
Bearing in mind  the bid-ask bounce effect (see for instance \cite{Andersen1999,Rhee1997}),
the factor $M_k$ in expression \eqref{tbtrmod} is equal to
\begin{equation}\label{mkbask}
M_k : = (-1)^{\xi_k} ~,
\end{equation}
where
\begin{equation}
 \xi_k : = \sum_{s=1}^k \iota_s ~.
 \end{equation}
The sequence $\{\iota_s\}$ takes into account the buy--sell structure of subsequent trades (ticks)
in the following sense: if the $(s-1)$-th and $s$-th trades are the same (i.e. both are selling or buying),
then $\iota_s=0$; in the opposite cases, $\iota_s=1$. We assume in our model that the
sequence $\{\iota_s\}$ consists of iid random Bernoulli integers, equal to one with probability $p$
and zero with the complementary probability $q= 1-p$.
The probability $q$ can be interpreted as the measure of distortion with respect to
the periodic reference buy-sell process. Indeed,
the closer $q$ is to zero, the more likely a buy (sell) trade
will be followed by a sell (buy) order. At $q=0$, the sequences $\{M_k\}$ and $\{r_k\}$
have their signs being exact periodic functions of $k$, reflecting a perfect
periodic buy-sell process that mimics a perfect bid-ask bounce.

\begin{definition}
For short, we refer to the parameter $q$ as the \emph{bounce distortion probability}.
\end{definition}

The positive iid denominators $\{H_k\}$ in expression \eqref{tbtrmod} are chosen
in such a way that the tick-by-tick returns $\{r_k\}$ possess a fat tail distribution
similar to that found in empirical data, as specified in Appendix A.3.

Our model assumes that the sequences $\{X_k\}$, $\{M_k\}$ and $\{H_k\}$ are mutually statistically independent. This implies in particular that the mean values of
the tick-by-tick returns $\{r_k\}$ are all equal to zero. It is easy to generalize
the theory developed below in the case where the means of $\{r_k\}$ are non-zero.

\subsection{Key properties}

In Appendix~A.1, we show that the correlations $\{\rho_m\}$
defined by \eqref{rhmdef} of the ARFIMA process $\{X_k\}$ are described
with excellent accuracy by the power law
\begin{equation}\label{rhompower}
\rho_m :=\text{E}\left[X_k X_{k+m}\right] \simeq
\begin{cases}
1 , & m = 0
\\
\digamma(\alpha) ~ m^{-\alpha} , & m \geqslant 1 ,
\end{cases}
\qquad \digamma(\alpha): = \frac{\Gamma(\frac{1+\alpha}{2})}{\Gamma(\frac{1-\alpha}{2})} ,
\end{equation}
where $\alpha =1-2d\in(0,1)$.

Appendix~A.2 derives that the correlation coefficients
of the bounce factors $\{M_k\}$ are given by
\begin{equation}\label{cmcorrgen}
\begin{array}{c} \displaystyle
C_m : = \text{E}\left[M_k~ M_{k+m}\right]  = e^{-\tilde{q}~ m} \times
\begin{cases}
(-1)^m , &  0< q < 1/2 ,
\\
1 , & 1/2 < q < 1 ,
\end{cases}
\\[6mm] \displaystyle
\qquad \tilde{q} := \ln\left(\frac{1}{|2 q-1|}\right) .
\end{array}
\end{equation}
The critical value $q_c=1/2$ divides the interval $q\in(0,1)$ into two parts, in which
the behavior of the correlation coefficients $C_m$ \eqref{cmcorrgen} are qualitatively different.
\begin{enumerate}
\item For $q\in(0,1/2)$, $C_m$ is a sign-alternating function of $m$.

\item For $q\in(1/2,1)$, $C_m$ is a positive non-oscillating function of its argument $m$.
\end{enumerate}
The existence of a sign-alternating behavior of $C_m$ for small $q$'s can be expected
from our previous remark that the closer $q$ is to zero, the more likely a buy (sell) trade
will be followed by a sell (buy) order.
Note also that, the closer $q$ is to the critical value $q_c=1/2$,
the more short-ranged becomes $C_m$. Below, we discuss only the case $q\in(0,1/2)$,
which is consistent with the known stylized fact that subsequent tick-by-tick returns
$r_k$ and $r_{k+ 1}$ are negatively correlated
(see for instance \cite{Andersen1999,Bouchaud2006,Cont2000}).

Appendix~A.3 describes the statistical properties that the
iid sequence $\{H_k\}$ need to obey in order for the
distribution $\phi(r)$ of tick-by-tick returns $\{r_k\}$ to be fat tailed.
In the following, we suppose for definiteness that $\phi(r)$ is a Student distribution of the form
\begin{equation}\label{pdfstudmu}
\phi(r) = \frac{\Gamma\left(\frac{\mu+1}{2}\right)}{b\sqrt{\pi} \, \Gamma\left(\frac{\mu}{2}\right)} \left(1+\frac{r^2}{b^2}\right)^{-\left(\frac{\mu+1}{2}\right)} ~,
\end{equation}
where corresponds to a tail exponent $\mu$.

\subsection{Correlation of tick-by-tick returns}

Putting all the properties described in the previous subsection together,
the correlation function of the tick-by-tick returns $\{r_k\}$ defined by \eqref{tbtrmod}
\begin{equation}\label{mathkmdef}
\mathcal{K}_m := \text{E}\left[r_k r_{k+m}\right] = \text{E}\left[X_k X_{k+m} M_k M_{k+m} \big/ H_k H_{k+m}\right]~,
\end{equation}
can be obtained explicitly.  Taking into account the mutual statistical independence of
the factors $\{X_k\}$, $\{M_k\}$ and
of the iid denominators $\{H_k\}$, and using relations, \eqref{rhompower}, \eqref{cmcorrgen}, one has
\begin{equation}\label{rkcorfuncall}
\begin{array}{c} \displaystyle
\mathcal{K}_m =  \varepsilon_2 ~ \mathcal{B}_m ~,
\\[4mm] \displaystyle
\mathcal{B}_m := e^{-\tilde{q}~ m}
\begin{cases}
1 , & m= 0 ~,
\\
\gamma \, (-1)^m m^{-\alpha} , & m \geqslant 1 ,
\end{cases}
\qquad \gamma := \digamma(\alpha) \frac{\varepsilon^2_1}{\varepsilon_2}~,
\end{array}
\end{equation}
where $\varepsilon_\theta:=\text{E}\left[H_k^{-\theta}\right]$  is
the inverse moment of the denominators $\{H_k\}$ given by expression \eqref{barrthetaexpr}.

Figure~1 shows the
autocorrelation function $\mathcal{K}_m/\varepsilon_2$ for $\alpha=0.1$ and for different values of
the bounce distortion probability $q$.

\subsection{Long-range correlations of absolute returns}

Consider now the autocorrelation function
\begin{equation}\label{absretcordef}
\mathcal{A}_m(\theta) : = \frac{\text{E}\left[|r_k|^\theta \, |r_{k+m}|^\theta \right] - \text{E}^2\left[|r_k|^\theta\right]}{\text{E}\left[|r_k|^{2\theta}\right]- \text{E}^2\left[|r_k|^\theta\right]}
\end{equation}
of power $\theta$ of the absolute values of the tick-by-tick returns.

Using the quadratic approximation \eqref{gthetrhosq} and relations \eqref{barrthetaexpr}, \eqref{rthdefs},
we can rewrite expression \eqref{matcalfnorm} in the form
\begin{equation}\label{matcalfnormsquara}
\mathcal{A}_m(\theta) \simeq
\begin{cases}
1 , & m = 0 ,
\\[4mm] \displaystyle
\chi \, m^{-\sigma} , & m\geqslant 1 ,
\end{cases}
\qquad \sigma := 2\alpha \in (0,2) ,
\end{equation}
where the factor $\chi$ is given by relation \eqref{betgam}.
It follows from the analysis of Appendix~A.4 that the power
law in (\ref{matcalfnormsquara}) constitutes
an extremely accurate description of $\mathcal{A}_m(\theta)$  for
a wide range of the parameters $\mu$, $\theta$ and for any $\sigma\in(0,2)$.
This is visualized in the log-log representation of figure~2, which shows the exact dependence of the
autocorrelation function of the absolute returns.

The main properties the correlation function \eqref{matcalfnormsquara} of the absolute returns are
as follows.
\begin{enumerate}
\item Relation \eqref{matcalfnormsquara} means that the correlations of
the absolute returns are long-ranged if the exponent $\sigma$ is sufficiently small (less than $1$
in order to technically qualify as ``long memory''  \cite{Beran}). Empirically, typical values of the exponent
are found in the range $\sigma\in (0.2,0.4)$ \cite{Cont2000}.

\item The exponent $\sigma$ is an empirically observable parameter as it can
be easily obtained from an appropriate statistical analysis of financial data.
This is in contrast with the exponent $\alpha$ of the correlation of returns given by \eqref{rkcorfuncall},
which is unobservable due to the distortion of the bid-ask bounce that
remove the information on this underlying power decay.

\item As shown in Appendix~A.4, the power law decay of the correlation of the absolute returns has the same tail exponent $\sigma$ for any $\theta\in(0.5,1.5)$, for which the quadratic approximation \eqref{gthetrhosq} of
the function $\widetilde{\mathcal{F}}(\theta,\rho)$ \eqref{ftildef} is highly accurate.

  \item The shape of the correlation function $\mathcal{A}_m(\theta)$ as a function of $\theta$ does not depend
on $m$ and $\sigma$ but is different for different values of the exponent $\mu$.
The dependences on $\theta$ of the normalized autocorrelation function
\begin{equation}\label{mathdmaxratio}
\Lambda(\theta) := \frac{\mathcal{A}_m(\theta)}{\max_{\,\theta} \mathcal{A}_m(\theta)} , \qquad m> 0~ ,
\end{equation}
for different values of the exponent $\mu$
of the Student distribution $\phi(r)$ \eqref{pdfstudmu}, are shown in figure~3.
We observe a shape similar to the daily returns correlations shown in Ref.~\cite{Ding1993}.
\end{enumerate}

\section{Correlation of returns in calendar time scale}

In the previous sections, we have described the statistical properties of returns
of our model in the tick-by-tick time scale. Henceforth, we
present a study of the statistical properties of instantaneous returns $R(t)$ \eqref{returnsdef}
and of $\Delta$-scale returns $R_\Delta(t)$ \eqref{rtdeldef} in calendar time scale.
This allows us to describe the microstructure noise and Epps effects.

\subsection{Correlation function of instantaneous returns}

The covariation function of instantaneous returns $R(t)$ \eqref{returnsdef} in the
calendar time scale representation defined by
$K(\tau) : = \text{E}\left[R(t) R(t+\tau)\right]$ is
obtained from the autocorrelation function $\mathcal{B}_m$ \eqref{rkcorfuncall}
at the tick-by-tick time scale by relation
\begin{equation}\label{natcovthruaucorm}
K(\tau) = \nu  \, \varepsilon_2 \left[ \delta(\tau) + B(\tau) \right]~,
\end{equation}
where
\begin{equation}\label{btaugenexpr}
B(\tau) = \sum_{m=1}^\infty \mathcal{B}_m f_m(|\tau|) ~,
\end{equation}
$\varepsilon_2$ is the variance of the tick-by-tick returns, $\nu$
is the mean rate of returns given by
\begin{equation}
\nu = 1/\bar{\tau} , \qquad \bar{\tau}:= \text{E}\left[\tau_{k,1}\right] ~,
\end{equation}
 while $f_m(\tau)$ ($\tau>0$) is the probability density function of tick-by-tick time interval durations $\tau_{k,m}$ \eqref{taukmdef}. A short outline of the derivation of
 relation \eqref{btaugenexpr} is given in Appendix~B.1.
It is convenient to use the mean time duration $\bar{\tau}$ between subsequent trades
as the unit time to scale all calendar times.

\begin{definition}
In the following, we scale all time scales by the  intertrade mean duration $\bar{\tau}$, referring to
this transformation as the ``calendar time scale representation''.
\end{definition}

In the calendar time scale representation, we have $\bar{\tau}\equiv 1$ and relation \eqref{natcovthruaucorm} takes the reduced form
\begin{equation}\label{natcovunittau}
K(\tau) = \varepsilon_2 \left[ \delta(\tau) + B(\tau) \right] .
\end{equation}

\subsection{Spectrum of instantaneous returns}

In what follows, we need to evaluate the spectrum of function $K(\tau)$ \eqref{natcovunittau}
\begin{equation}
\tilde{K}(\omega) : = \frac{\varepsilon_2}{\pi} \int_0^\infty K(\tau) \cos(\omega\tau) d\tau~ .
\end{equation}
Using \eqref{btaugenexpr}, \eqref{natcovunittau}, we obtain
\begin{equation}\label{atilones}
\tilde{K}(\omega) = \frac{\varepsilon_2}{2\pi} \left[ 1 + \tilde{B}(\omega) \right] ,
\end{equation}
where
\begin{equation}\label{somgendef}
\tilde{B}(\omega) : = 2 \int_0^\infty B(\tau) \cos(\omega \tau) d\tau =  2 \text{Re}\left[ \sum_{m=1}^\infty \mathcal{K}_m \hat{f}_m(i \omega) \right]
\end{equation}
and $\hat{f}_m(s)$ is the Laplace image of the probability density function $f_m(\tau)$
\begin{equation}
\hat{f}_m(s) : = \int_0^\infty f_m(\tau) e^{-s \tau} d \tau .
\end{equation}
Although some authors are found evidence of long-ranged correlations between different inter-trade time intervals \cite{Ivanov2004}, we, along with Ref.~\cite{Scalas2007}, suggest henceforth that tick-by-tick instants $\{t_k\}$ are such that the random inter-trade durations $\{\tau_k\}$ are iid random variables with the same
probability density function $f(\tau)$.

\begin{assertion}
Our microstructure return model  \eqref{tbtrmod} of tick-by-tick returns is fully
defined when supplemented by the specification of the iid sequence $\{\tau_k\}$ of inter-trade intervals
distributed according to the probability density function $f(\tau)$.
\end{assertion}

Within our microstructure return model, the probability density functions $f_m(\tau)$ are equal to
m-tiple convolutions of $f(\tau)$. Accordingly, relation \eqref{somgendef} takes the form
\begin{equation}
\tilde{B}(\omega) : = 2 \text{Re}\left[ \sum_{m=1}^\infty \mathcal{K}_m \hat{f}^m(i \omega) \right] .
\end{equation}
Substituting the expression \eqref{rkcorfuncall} for the autocorrelation function of
the tick-by-tick returns and after summing the series, we obtain
\begin{equation}\label{somintrepr}
\tilde{B}(\omega) = - \frac{2 (1-2 q) \gamma}{\Gamma(1+\alpha)} \int_0^\infty \text{Re}\left[ \frac{\hat{f}(i\omega)}{e^{u^{1/\alpha}} +(1-2 q)  \hat{f}(i\omega)} \right] du .
\end{equation}
The factor $\gamma$ is defined from \eqref{rkcorfuncall} with \eqref{rhompower}.

\subsection{Statistics of inter-trade time intervals}

In what follows, we use relation \eqref{somintrepr}
to analyze quantitatively the microstructure noise and Epps effects.
The characteristics of these effects depend essentially on the
statistics of the inter-trade time intervals. Under
the condition that the inter-trade durations are iid, all information about their statistics
is contained in the probability density function $f(\tau)$.
We thus first discuss possible models for this probability density function.

In  the econometric literature, the statistics of
inter-trade intervals has been studied empirically and parametric models
have been proposed (see, for instance, \cite{Ivanov2004,Politi2008,Sazuka2007}).
Here, we discuss a few analytical models for
the probability density functions $f(\tau)$, which provide reasonable fits
to the empirical distributions of inter-trade intervals while at same time
possess a convenient analytical expression for their Laplace images $\hat{f}(s)$.
Having such analytical approximations is very useful for the determination
of the spectrum $\tilde{B}(\omega)$ \eqref{somintrepr}, which is defined
in terms of the Laplace image $\hat{f}(i \omega)$ of the probability density function of
inter-trade intervals expressed in terms of the imaginary argument $i \omega$.

A model often used to represent the distribution of inter-trade intervals is the Weibull distribution
\begin{equation}
f(\tau) = \lambda w(\lambda\tau;\beta) , \qquad w(\tau;\beta) := \beta \tau^{\beta-1} e^{-\tau^\beta}~ ,
\end{equation}
where $\beta$ is the shape parameter and $\lambda$ is the scaling parameter.
The corresponding survival probability $Q(\tau) := \int_\tau^\infty f(u) du$
is given by expression
\begin{equation}\label{cweibdef}
Q(\tau) = W(\lambda\tau;\beta) , \qquad
W(\tau;\beta) := \int_\tau^\infty w(u;\beta)\, du = e^{-\tau^\beta} .
\end{equation}
In the reduced time scale representation in which the unit duration is taken as the
mean inter-trade interval $\bar{\tau}$, the Weibull distribution
depends only on the shape parameter $\beta$, while the scale parameter $\lambda$
is defined by the condition
\begin{equation}\label{meannormcond}
\bar{\tau} = \int_0^\infty Q(\tau)d\tau \equiv 1 \qquad \Rightarrow \qquad \lambda = \lambda_w(\beta) := \Gamma\left(\frac{1+\beta}{\beta}\right) .
\end{equation}

Figure~4 presents typical time sequences $\{t_k\}$ of trades with $t_k = \sum_{i=1}^k \tau_i$,
where the inter-trade intervals $\tau_i$'s are drawn from the
Weibull distribution with shape parameters $\beta=0.5; 1; 2$ respectively.
Notwithstanding the mutual independence of the
inter-trade intervals $\{\tau_k\}$, one can observe clustering for $\beta<1$ and
quasi-periodicity for $\beta>1$.

However, the Weibull distribution family has some shortcoming.
First, its Laplace image does not have a simple analytical representation for any value of $\beta$.
Second, it has not enough shape parameters in order to fit the
empirical probability density function over the whole range of $\tau$ values
for which the empirical probability density function is still measured accurately.
Indeed, using $\beta \simeq 0.75\div 0.85$, the Weibull distribution provides a
rather accurate representation of the empirical distribution at small and intermediate values of
$\tau$. But it goes to zero too fast for large $\tau$ values, compared with the
slow decay of the empirical probability density function \cite{Politi2008}.
This justifies using some more flexible distributions, which are related to the Weibull
family. One such possibility is the generalized gamma distribution (GGD)
\begin{equation}\label{gengamdistlamdef}
f(\tau) = \lambda g(\lambda\tau;\vartheta,\beta) , \qquad g(\tau;\vartheta,\beta) := \frac{\beta \tau^{\vartheta-1}}{\Gamma(\vartheta/\beta)} \, e^{-\tau^\beta}~ .
\end{equation}
Its survival probability is
\begin{equation}\label{weibsurvgendef}
Q(\tau) = G(\lambda \tau;\vartheta,\beta) , \qquad
G(\tau;\vartheta,\beta) :=\frac{\Gamma\left(\vartheta/\beta, \tau^\beta \right)}{\Gamma\left(\vartheta/\beta\right)}~,
\end{equation}
where $\Gamma(a,z)$ is the incomplete gamma function.
The scale parameter $\lambda$ is given by
\begin{equation}
\lambda = \lambda_g(\vartheta,\beta) : = \frac{\Gamma[(1+\vartheta)/\beta]}{\Gamma(\vartheta/\beta)}
\end{equation}
so as to ensure that the mean time duration between trades is unity.

From the point of view of the feasibility of analytical calculations of
the integral \eqref{somintrepr}, the GGD has the nice property of having
analytical expressions for its Laplace images for any rational $\beta$
and for arbitrary shape parameter $\vartheta$. Analytical expressions of
the Laplace images of the GGD \eqref{gengamdistlamdef}, for some rational values of
it shape parameter $\beta$, are given in Appendix~B.2.

Figure~5 shows the survival functions of the Weibull distribution \eqref{cweibdef} for $\beta=0.8$
and of the generalized gamma distribution \eqref{weibsurvgendef}, for the same $\vartheta=0.8$ and for $\beta=2/3$, $1/2$ and $\beta=1/3$.
Figure~6 compares the function $\tilde{B}(\omega)$ \eqref{somintrepr}
obtained for a Poissonian statistics of the tick-by-tick times $\{t_k\}$, where
\begin{equation}\label{pdflapois}
f(\tau) = e^{-\tau} \qquad \Leftrightarrow \qquad \hat{f}(s) = \frac{1}{1+s}~ ,
\end{equation}
to the two functions $\tilde{B}(\omega)$ for the GGDs for $\theta=0.8$ and $\beta=1/2; 2/3$, for which
the Laplace images of the probability density function are given by the analytical expressions \eqref{hatonttwoexpl}, \eqref{gengamonetwolap} and \eqref{thethreelam}, \eqref{twothree}. 

All curves represented in figure~6 coincide at $\omega=0$, which is the consequence of the fact that $\tilde{B}(0)$ does not depend on the shape of
the distribution $f(\tau)$ of the inter-trade interval durations. Note however the significant differences
between the three realizations of $\tilde{B}(\omega)$ at $\omega\neq 0$, which emphasizes
the difference between the exponential probability density function \eqref{pdflapois}
and the GGD.

\subsection{Short-ranged correlations of discrete returns in calendar time}

In section~2, we have demonstrated that our microstructure return model
predicts that the range of the correlations of the tick-by-tick returns
is controlled essentially by the bounce distortion probability $q$ (see figure~1).
 In particular, if $q$ is close to zero, then the autocorrelation $\mathcal{K}_m$ \eqref{rkcorfuncall}
 of the tick-by-tick returns is a long-ranged (sign-alternating) function of $m$.
 In contrast, if $q$ is close to the critical value $q_0=1/2$, the
 correlation of the tick-by-tick returns becomes short ranged (as an illustration,
 see the upper and lower plots in figure~1).
We now show that, even for small $q$'s including $q=0$, the
correlation of the $\Delta$-scale returns $R_\Delta(t)$ \eqref{rtdeldef} in
calendar time is short-ranged, as a result of the washing action of the
random inter-trade intervals $\{\tau_k\}$.

To demonstrate this result, we rewrite the $\Delta$-scale returns in the more convenient form
\begin{equation}\label{rdeltdef}
R_\Delta(t) = \Pi_\Delta(t) \otimes R(t)  , \qquad \Pi_\Delta(t) : =
\begin{cases}
1 , & t\in (0, \Delta) ,
\\
0 , & t \notin (0,\Delta) ,
\end{cases}
\end{equation}
where the symbol $\otimes$ represents the convolution operation.
Using \eqref{natcovunittau} and \eqref{rdeltdef},
the correlation function of the $\Delta$-scale returns $R_\Delta(t)$ is given by the convolution
\begin{equation}\label{adelcorconv}
\begin{array}{c}
K_\Delta(\tau) :=  \text{E}\left[R_\Delta(t) R_\Delta(t+\theta)\right] = K(\tau) \otimes \mathcal{T}_\Delta(\tau) =
\\[4mm] \displaystyle
\varepsilon_2  \left[ \delta(\tau) + B(\tau) \right]  \otimes \mathcal{T}_\Delta(\tau) ~,
\end{array}
\end{equation}
where
\begin{equation}\label{mathtdeldef}
\mathcal{T}_\Delta(\tau) :=
\begin{cases}
\Delta - |\tau| , & |\theta|<\Delta ,
\\
0 , & |\tau|>\Delta .
\end{cases}
\end{equation}

For the numerical estimation of the function $K_\Delta(\tau)$, we use relation
\eqref{adelexpr}, which is directly derived from relation \eqref{adelcorconv}.
Figure~7 shows the ratio $K_\Delta(\tau)/K_\Delta(0)$, for $\Delta=1$ and for different values of the bounce distortion probability $q$. One can observe the short range nature of the
correlations of the $\Delta$-scale returns in calendar time for all $q$ values.
Note the fact that the correlations become negative before decaying to zero.
The results shown in figure~7 use the GGD \eqref{gengamdistlamdef} for
the distribution $f(\tau)$ of the inter-trade intervals, with $\vartheta =0.8$ and $\beta=2/3$.
Analogous plots for $q=0$ and for different values of $\Delta$ are shown in figure~8.

\section{Microstructure noise effect}

\subsection{Basic notions}

The goal of this section is to show that our model provides a natural set-up
for the microstructure noise effect \cite{Ait2009,Andersen1999,Bacry2010,Bandi2006,Bandi2008,Cont2000}.
The microstructure noise effect refers to the following phenomenon.
Let us consider the realized volatility\footnote{In this context, volatility is defined as the variance of the log-price increments over a given time interval of duration $\Delta$.}, equal to
\begin{equation}\label{realvoldef}
\hat{D}(\Delta,T) = \frac{1}{T} \sum_{k=1}^{\lfloor T/\Delta \rfloor} R^2_\Delta(k\Delta)~ .
\end{equation}
If $R_\Delta(t)$ is a stationary and ergodic process, then the realized volatility \eqref{realvoldef} converges
in probability as $T$ goes to infinity to
\begin{equation}\label{volexpval}
D(\Delta) := \frac{1}{\Delta} \text{E}\left[R_\Delta^2(t)\right] ,
\end{equation}
which is, by definition, the average volatility density over intervals of duration $\Delta$.

The geometric Brownian motion (GBM)  $L(t) = D \cdot W(t)$ is the simplest and often used first-order model
of price dynamics, where $W(t)$ is the standard Wiener process. The GBM is such that
the volatility density $D(\Delta)$ does not depend on $\Delta$ ($D(\Delta)\equiv D$), so that
its estimation is in principle independent of the time durations $\Delta$. In addition, for GBM processes,
the volatility is  \emph{observable} as, for $\Delta\to 0$ and for any given observation interval
of duration $T=\text{const}$, the realized volatility converges in probability to the volatility density $D$:
\begin{equation}
\hat{D}(\Delta,T) ~ \overset{\text{P}}{\longrightarrow} ~ D , \qquad \Delta \to 0 , \qquad T=\text{const}~ .
\end{equation}

Real financial markets depart from the ideal GBM and exhibit the
\emph{microstructure noise effect}, i.e., for
small durations $\Delta$ that are comparable to the mean time interval $\bar{\tau}$ between
subsequent trades, the realized volatility is positively biased. Mathematically,
this means that $D(\Delta)$ defined by expression \eqref{volexpval}
increases as $\Delta$ decreases.

\begin{assertion}
The next subsection demonstrates that the microstructure noise
effect results simply from the negative return correlations inherent in the
microstructure model proposed in the present paper.
\end{assertion}

\subsection{Strength of microstructure noise effect}

In order to describe quantitatively the microstructure noise effect, let
us calculate the function $D(\Delta)$ defined by expression \eqref{volexpval}
in the frame of our microstructure returns model. Due to equality \eqref{adelcorconv}, one has
\begin{equation}\label{ddeltathrub}
D(\Delta) = \frac{1}{\Delta} K_\Delta(0) =  \varepsilon_2 \left( 1  +  \frac{2}{\Delta} \int_0^\infty \mathcal{T}_\Delta(\tau) B(\tau) d\tau \right)~ .
\end{equation}

\begin{definition}
We defined respectively by ``true'' and ``microstructure'' volatilities the following limit values
\begin{equation}\label{Dtrmicdefs}
D_\text{\normalfont{true}} := \lim_{\Delta\to\infty} D(\Delta)~ , \qquad
D_\text{\normalfont{micro}} := \lim_{\Delta\to 0} D(\Delta)~ .
\end{equation}
The ratio
\begin{equation}\label{strengthdef}
\mathcal{S} := \frac{D_\text{\normalfont{micro}}}{D_\text{\normalfont{true}}}
\end{equation}
defines the strength of the microstructure noise effect.
\end{definition}
The ratio $\mathcal{S}$ \eqref{strengthdef} has the following intuitive
economic meaning. It is equal to the dimensionless ratio of
the realized volatility $D(\Delta)$ \eqref{volexpval} at the micro ($\Delta\to 0$)
and at the macro ($\Delta\to\infty$) scales. This justifies that $\mathcal{S}$
provides a convenient quantitative characterization of the strength of the microstructure noise effect.

The dependence of the strength $\mathcal{S}$ \eqref{strengthdef} of the microstructure noise effect
with respect to the parameters of our microstructure noise model is obtained as follows.
From expressions \eqref{ddeltathrub}, \eqref{somintrepr} and \eqref{mathtdeldef}, we obtain
\begin{equation}\label{thruedexpr}
D_\text{\normalfont{micro}} = \varepsilon_2 , \qquad D_\text{true}= \varepsilon_2 [1+ \tilde{B}(0) ] ,
\end{equation}
so that,
\begin{equation}\label{msnstrengthdef}
\mathcal{S} = \frac{1}{1+\tilde{B}(0)} .
\end{equation}
In view of relation \eqref{somintrepr}, one has
\begin{equation}
\tilde{B}(0) = - \frac{2 (1-2 q) \gamma}{\Gamma(1+\alpha)} \int_0^\infty \frac{du }{e^{u^{1/\alpha}} + 1-2 q }~ .
\label{fgjyj}
\end{equation}
Expression \eqref{msnstrengthdef} with (\ref{fgjyj}) shows that $\mathcal{S}$ does not depend
on the shape of the probability density function $f(\tau)$ of the inter-trade interval durations.
It does depend on the other parameters $\alpha, q, \mu$ of the model, where
$\alpha$ is the unobservable exponent of the power law correlation \eqref{rhompower} of
the auxiliary ARFIMA process $\{X_k\}$, $q$ is the bounce distortion probability,
and  $\mu$ is the exponent of the power law describing the tail of
the probability density function $\phi(r)$ \eqref{pdfstudmu} of the tick-by-tick returns.
While $\alpha$ is not directly observable, it can be derived from the
the observable exponent $\sigma$ of the power law dependence of the
correlation \eqref{matcalfnormsquara} of the absolute tick-by-tick returns
via the equality $\sigma=2\alpha$.
Figures 9-11 illustrate the dependence of the strength $\mathcal{S}(\alpha, q, \mu)$ of
the microstructure noise effect on these three parameters.

A further quantification of the strength of the microstructure noise effect is obtained
by introducing the generalization of the strength $\mathcal{S}$ \eqref{strengthdef} defined by
\begin{equation}
\mathcal{S}_\Delta := \frac{D(\Delta)}{D_\text{true}} ~,
\label{fgjujueg}
\end{equation}
which now depends on the interval duration $\Delta$ over which the returns
$R_\Delta(t)$ are defined. Figure~12 illustrates the dependence of
$\mathcal{S}_\Delta$ for $\alpha=0.1$ ($\sigma=0.2$), $\mu=5$ and
for different values of the bounce distortion probability $q$.

\section{Epps effect}

\subsection{Epps effect: basic notions}

The microstructure return model introduced in the present paper
is able to identify the roots of the Epps effect \cite{Epps1979,Toth2009}, as we now demonstrate.

For this, let us describe the Epps effect by using the notations introduced above.
Let us consider the instantaneous returns $R_1(t)$ and $R_2(t)$ of two
assets. Analogously to \eqref{rdeltdef}, the $\Delta$-scale returns
of the two assets during  a time interval of duration $\Delta$
can be expressed with the relations
\begin{equation}
R_{1,\Delta}(t) = \Pi_\Delta(t) \otimes R_1(t), \qquad R_{2,\Delta}(t) = \Pi_\Delta(t) \otimes R_2(t) ~.
\end{equation}
The Epps effect is observed on a quantity that is
analogous to \eqref{volexpval}, except that under the
expectation sign the square return $R^2_\Delta(t)$ is replaced
by the product of the two different returns $R_{1,\Delta}(t)$ and $R_{2,\Delta}(t)$
\begin{equation}\label{crossreturns}
D_{1,2}(\Delta) := \frac{1}{\Delta} \text{E}\left[R_{1,\Delta}(t) R_{2,\Delta}(t)\right] ~.
\end{equation}
$D_{1,2}(\Delta)$ is thus a measure of inter-dependence between the two assets.

It is convenient to introduce a normalized version
of $D_{1,2}(\Delta)$ \eqref{crossreturns}. For this, we consider the simple
case where the returns $R_1(t)$ and $R_2(t)$ are statistically equivalent in the sense that
$D_1(\Delta) = D_2(\Delta) = D(\Delta)$, where
\begin{equation}
D_1(\Delta) := \frac{1}{\Delta} \text{E}\left[R_{1,\Delta}^2(t)\right] , \qquad D_2(\Delta) := \frac{1}{\Delta} \text{E}\left[R_{2,\Delta}^2(t)\right]~ .
\end{equation}
The normalized version of $D_{1,2}(\Delta)$ \eqref{crossreturns} is defined as
\begin{equation}\label{rhocross}
\mathcal{S}_\Delta^{1,2} := \frac{D_{1,2}(\Delta)}{D_\text{true}} , \qquad D_\text{true}:=\lim_{\Delta\to\infty} D(\Delta) .
\end{equation}
Now we are ready to formulate the Epps effect.

\begin{definition}
\textnormal{\it Consider two assets whose $\Delta$-scale returns $R_{1,\Delta}(t)$ and $R_{2,\Delta}(t)$
at scale $\Delta$ are correlated so that the following limit is positive
\begin{equation}
\lim_{\Delta\to\infty} \mathcal{S}_\Delta^{1,2} > 0 ~.
\end{equation}
The Epps effect corresponds to the fact that
$\mathcal{S}_\Delta^{1,2}$ \eqref{rhocross} is a monotonically increasing function of
the argument $\Delta$, and which vanishes as $\Delta \to 0$.}
\end{definition}

\subsection{Epps effect paradox}

Before explaining the Epps effect in the frame of our
microstructure returns model, it is illuminating to
first discuss a paradox that emerges when comparing the dependence with $\Delta$
of the two similar functions
\begin{equation}\label{dparadox}
D(\Delta) = \frac{1}{\Delta} \text{E}\left[R_{1,\Delta}^2(t)\right] \quad \text{and} \quad D_{1,2}(\Delta) = \frac{1}{\Delta} \text{E}\left[R_{1,\Delta}(t) R_{2,\Delta}(t)\right] ~.
\end{equation}
According to the microstructure noise effect discussed in the previous section,
the first $D(\Delta)$ is a \emph{decreasing} function of $\Delta$ while, according to the Epps effect,
the second $D_{1,2}(\Delta)$ is an \emph{increasing} function of $\Delta$.
The paradox is that, while $R_{2,\Delta}(t)$ may ``only insignificantly'' differ from $R_{1,\Delta}(t)$,
replacing one $R_{1,\Delta}(t)$ in the definition of $D(\Delta)$ by $R_{2,\Delta}(t)$,
i.e., changing $R_{1,\Delta}^2(t)$ into  $R_{1,\Delta}(t) R_{2,\Delta}(t)$, change an increasing
function into a decreasing function of $\Delta$.

In order to elucidate this paradoxical bifurcation of the function $D(\Delta)$ under
the change of $R_{1,\Delta}^2(t)$ into  $R_{1,\Delta}(t) R_{2,\Delta}(t)$,
let us consider, analogously to \eqref{returnsdef}, the instantaneous returns
\begin{equation}
R_1(t) = \sum_{k} r_{1,k} ~ \delta(t-t_{1,k}) ~.
\end{equation}
This defines a singular stochastic process with infinite mean:
\begin{equation}\label{ersqinfty}
\text{E}\left[R_1^2(t)\right] = \infty .
\end{equation}
Consider now another instantaneous returns process
\begin{equation}
R_2(t) = \sum_{k} r_{2,k} ~ \delta(t-t_{2,k})~ ,
\end{equation}
and let us assume that it is only infinitesimally different from $R_1(t)$, by
assuming that both tick-by-tick returns are identical
($r_{2,k} \equiv r_{1,k} ,  \forall ~ k$), while
the instants $\{t_{2,k}\}$ of the second return process
are only infinitesimally different from the instants $\{t_{1,k}\}$ of the first return process:
\begin{equation}
t_{2,k} = t_{1,k} + \zeta , \qquad \forall ~ k ~.
\end{equation}
The time shift $\zeta\neq 0$ is supposed to be infinitesimal so that we can write
\begin{equation}\label{r2istretshift}
R_2(t) = \sum_{k} r_k ~ \delta(t-t_k-\zeta) = R_1(t-\zeta) ~,
\end{equation}
together with
\begin{equation}\label{r1istretshift}
R_1(t) = \sum_{k} r_k ~ \delta(t-t_k) , \qquad t_{k} := t_{1,k} , \qquad r_{k} := r_{1,k} = r_{2,k}~.
\end{equation}
In this case, contrary to equality \eqref{ersqinfty}, and notwithstanding the ``infinitesimal difference'' between $R_1(t)$ and $R_2(t)$, the following equality holds
\begin{equation}
\text{E}\left[R_1(t)R_2(t)\right] \equiv 0~ .
\end{equation}
This suggests that the resolution of the paradox rests on the fact that the so-called
infinitesimal difference or shift of the tick-by-tick instants drastically changes
the correlation between the different instantaneous returns, ``reducing infinity to zero''.

The same conclusion holds qualitatively for $\Delta$-scale returns defined over a finite time interval $\Delta$.
Indeed, let us go from the instantaneous singular returns \eqref{r1istretshift}, \eqref{r2istretshift}
to the regular returns
\begin{equation}\label{r12retshift}
R_{1,\Delta}(t) = \sum_{k} r_k ~ \Pi_\Delta(t-t_k) , \qquad R_{2,\Delta}(t) = \sum_{k} r_k ~ \Pi_\Delta(t-t_k-\zeta) ,
\end{equation}
where $\zeta$ is a nonzero shift. Figure~13 shows a realizations of these two $\Delta$-scale return processes
for $\zeta \gtrsim \Delta$, where both $\zeta$ and $\Delta$ are significantly smaller than mean inter-trade time duration $\bar{\tau}$. Although the $\Delta$-scale returns $R_{1,\Delta}(t)$ and $R_{2,\Delta}(t)$ are almost identical, their product is equal to zero
\begin{equation}\label{r12zeroident}
R_{1,\Delta}(t) \cdot R_{2,\Delta}(t) = 0~ ,
\end{equation}
which is in accordance with the Epps effect, suggesting that the function $D_{1,2}(\Delta)$ \eqref{crossreturns} is vanishing at $\Delta\to 0$.

Figure~14 shows another realization of the two $\Delta$-scale return processes
\eqref{r12retshift} now for $\zeta\lesssim \Delta$, which implies that
\begin{equation}
R_{1,\Delta}(t) \cdot R_{2,\Delta}(t) \equiv R_{2,\Delta-\zeta}^2(t) \neq 0~ .
\end{equation}
Accordingly, as can  be seen in figure~14, the function $D_{1,2}(\Delta)$ \eqref{crossreturns} is equal to
\begin{equation}\label{dellargepsepps}
\begin{array}{c} \displaystyle
D_{1,2}(\Delta) \simeq \frac{1}{\Delta} \text{E}\left[R_{\Delta-\zeta}^2(t)\right] \simeq \nu \varepsilon_2 ~ \frac{\Delta-\zeta}{\Delta},
\\[4mm] \displaystyle
\Delta > \zeta , \qquad \Delta \ll \bar{\tau} ,
\end{array}
\end{equation}
and increases (for $\Delta>\zeta$) with $\Delta$ increasing, in accordance with Epps effect.

\begin{assertion}
In sum, the Epps effect is stipulated by the (random) shifts between respective trades instants $t_{1,k}$ and $t_{2,k}$ of different assets. Due to these shifts, the smaller the scale $\Delta$ of returns $R_{1,\Delta}(t)$ and $R_{2,\Delta}(t)$, the less likely it is that trade instants $t_{1,k}$ and $t_{2,k}$ belong to the same time interval $(t-\Delta,t)$. As a result, the $\Delta$-scale returns $R_{1,\Delta}(t)$ and $R_{2,\Delta}(t)$ become less correlated, thus giving rise to the Epps effect. The quantitative description of the Epps effect in the framework of our model of returns is considered below.
\end{assertion}

\subsection{Epps effect: description in the frame of microstructure returns model}

To quantify the Epps effect in our microstructure returns model, we
calculate the two functions $D_{1,2}(\Delta)$ \eqref{crossreturns} and
$\mathcal{S}_\Delta^{1,2}$ \eqref{rhocross}. For this, we need to estimate the
cross correlations of the instantaneous returns
\begin{equation}
K_{1,2}(\tau):= \text{E}\left[R_1(t) R_2(t+\tau)\right] ~.
\end{equation}
We consider the simplest model representing the inter-dependence between
$R_1(t)$ and $R_2(t)$, which reads
\begin{equation}\label{r2delaydef}
R_1(t) :  = R(t) , \qquad R_2(t) := R(t+\zeta)~ ,
\end{equation}
where $R(t)$ is some instantaneous return process and $\zeta$ is some random delay time
distributed according to some known
probability density function $\kappa(\tau)$. If $\zeta$ is always positive,
$R_2(t)$ can be said to be \emph{subordinated} to $R_1(t)$ but we do not specifically
need this to hold true. We do impose however that the inter-dependence between
the two return processes is symmetric, which is reflected into the evenness of the
probability density function: $\kappa(-\tau)=\kappa(\tau)$.

Using relations \eqref{r2delaydef} and  equality \eqref{natcovunittau} for
the correlation function $K(\tau)$, we obtain
\begin{equation}
K_{1,2}(\tau) := K(\tau)\otimes \kappa(\tau) = \varepsilon_2 \left[ \kappa(\tau) +  B(\tau)\otimes \kappa(\tau) \right] ~ .
\end{equation}
We can then represent the function $D_{1,2}(\Delta)$ \eqref{crossreturns} in
a form analogous to relation \eqref{ddeltathrub}:
\begin{equation}\label{d12timeint}
D_{1,2}(\Delta) = \frac{2 \varepsilon_2}{\Delta}
\int_{0}^\infty \left[ \kappa(\tau) +  B(\tau)\otimes \kappa(\tau) \right] \mathcal{T}_\Delta(\tau) d\tau~ .
\end{equation}

To make the analytical calculations explicit, we assume for
definiteness that the probability density function
of the random delay time $\zeta$ is Gaussian with zero mean and variance $\lambda^2$.
In our numerical calculations, we use relation \eqref{donetwospectr}
which is equivalent to expression \eqref{d12timeint}.

Figure 15 shows the dependence of the
function $\mathcal{S}^{1,2}_\Delta$ defined in \eqref{rhocross} as a function of $\Delta$,
clearly demonstrating the existence of the Epps effect in the frame of our microstructure returns model.
We do not present plots for different $\alpha$ and $q$ values because
our calculations show that the dependence of $\mathcal{S}^{1,2}_\Delta$
as a function of $\Delta$ is practically undistinguishable
for different $\alpha$ values and for any $q>0$.

\pagebreak

\appendix
\setcounter{section}{0}
\setcounter{equation}{0}
\setcounter{theorem}{0}
\renewcommand{\theequation}{\thesection.{\arabic{equation}}}
\renewcommand{\thetheorem}{\thesection.{\arabic{theorem}}}
\renewcommand{\thesection}{\Alph{section}}

\section{Statistics of tick-by-tick returns}

\setcounter{equation}{0}
\setcounter{theorem}{0}
\setcounter{remark}{0}
\setcounter{lemma}{0}
\setcounter{definition}{0}
\setcounter{example}{0}
\setcounter{supposition}{0}

In this appendix, we give a detailed description of our tick-by-tick
microstructure return model and derive its basic statistical properties.

\subsection{Statistical properties of ARFIMA process}

We explore the statistical properties of the tick-by-tick returns $\{r_k\}$ \eqref{tbtrmod}, beginning with
a discussion of the correlation properties of the ARFIMA process $\{X_k\}$.

It is known that the solution of equation \eqref{fracxeq} is
\begin{equation}\label{xkthruukjsol}
X_k = \Upsilon \sum_{j=0}^\infty a_j u_{k-j} , \qquad a_j = \frac{\Gamma(j+d)}{\Gamma(j+1) \Gamma(d)} .
\end{equation}
Its correlation function is equal to
\begin{equation}
\rho_m : = \text{E}\left[X_k X_{k+m}\right] = \sum_{j=0}^\infty a_j a_{j+m} ,
\label{fjrukjik}
\end{equation}
where the $\{a_j\}$'s are defined in \eqref{xkthruukjsol}. Calculating
the sum in (\ref{fjrukjik}) and choosing the factor $\Upsilon$ in \eqref{xkthruukjsol} such that $\rho_0=1$,
we obtain
\begin{equation}\label{rhomexpr}
\rho_m = \frac{\Gamma(1-d)}{\Gamma(d)} \cdot \frac{\Gamma(d+m)}{\Gamma(1-d+m)} .
\end{equation}
Using the well-known asymptotical relation
\begin{equation}
\frac{\Gamma(d+m)}{\Gamma(1-d+m)} \simeq m^{-\alpha} , \qquad \alpha = 1-2d~, \qquad m \gg 1 ,
\label{fjuykiumn}
\end{equation}
we conclude that $\rho_m$ is asymptotically a power law.

For fractional orders $d\in(0,1/2)$ of the difference lag equation \eqref{fracxeq},
the expansion (\ref{fjuykiumn}) holds accurately even for small $m$ values
($m\gtrsim 1$). With good approximation, one may thus express $\rho_m$ \eqref{rhomexpr} by
\begin{equation}
\varrho_m :=
\begin{cases}
1 , & m = 0
\\
\digamma(\alpha) ~ m^{-\alpha} , & m \geqslant 1 ,
\end{cases}
\qquad \digamma(\alpha): = \frac{\Gamma(\frac{1+\alpha}{2})}{\Gamma(\frac{1-\alpha}{2})} ,
 \qquad \alpha = 1- 2 d .
\label{rhompower2}
\end{equation}
Figure~A1 illustrates the accuracy of the power law approximation (\ref{rhompower2})
transposed in the main text as expression \eqref{rhompower}. One can observe
a relative error of no more than 1.4\% even for $m=1$.

Moreover,  it follows from expressions \eqref{rhomexpr} that, for any $m\geqslant 1$,
\begin{equation}
\rho_m \leqslant \rho_1(d) : = \frac{\pi d}{\sin(\pi d)\, \Gamma^2(d)}~ .
\label{ehju6pjumj}
\end{equation}
Figure~A2 shows the dependence of
$\rho_1(d)$ as a function of $d$. For any $m\geqslant 1$ and $d\in(0,1/2)$,
the correlation function $\rho_m$ satisfies the inequality
\begin{equation}\label{rhomineq}
0\leqslant\rho_m \leqslant 1/2 , \qquad m\geqslant 1 , \qquad d\in (0,1/2) .
\end{equation}

\subsection{Correlation of the bid-ask bounce factor $M_k$}

The correlation of the tick-by-tick returns $\{r_k\}$ is significantly
influenced by the correlation $C_m : = \text{E}\left[M_k~ M_{k+m}\right]$ of the bid-ask bounce factor $M_k$
defined by expression \eqref{mkbask}.
Using the fact that
\begin{equation}
(-1)^{\xi_k+\xi_{k+m}} \equiv (-1)^{\delta_\xi(k,m)} , \qquad \delta_\xi(k,m) : = \xi_{k+m} - \xi_k = \sum_{s=k}^{k+m} \iota_s ~,
\end{equation}
we obtain
\begin{equation}
C_m = \text{E}\left[(-1)^{\delta_\xi(k,m)}\right] ~.
\end{equation}
Assuming that the sequence $\{\iota_s\}$ consists of iid random Bernoulli integers, equal to one with probability $p$ and zero with probability $q= 1-p$, this implies that $\delta_\xi(k,m)$ is, for a given $m$,
a Binomial variable such that
\begin{equation}
\Pr\{\delta_\xi(k,m)=s\} = \binom{m}{s} p^s q^{m-s} , \qquad s= 0, 1, \dots , m .
\end{equation}
Accordingly, the correlation $C_m$ is equal to
\begin{equation}
C_m = (2 q-1)^m~,
\end{equation}
which may be rewritten in the form \eqref{cmcorrgen}.

\subsection{Statistics of the denominator $H_k$ in expression  \eqref{tbtrmod}}

The iid positive random variables$\{H_k\}$ have been introduced in the definition of our
microstructure return process in order to
adjust the statistical properties of the tick-by-tick returns \eqref{tbtrmod}
to those of the empirical tick-by-tick returns.

In order to determine what should be their distribution $\varphi(\eta)$, we rewrite
equality  \eqref{tbtrmod} in the more convenient for our subsequent analysis:
\begin{equation}\label{rkeqykhk}
r_k = Y_k / H_k , \qquad Y_k = X_k M_k .
\end{equation}
From the structure of $M_k$'s, we conclude that
the probability density function of each numerator $Y_k$ is, like $X_k$, possess by Gaussian probability density function $\psi(y)$ with zero mean and unit variance.

We specify the probability density function $\varphi(\eta)$ of the $H_k$'s
so that the distribution of tick-by-tick returns is a fat-tail distribution similar to the
empirical returns (see, for instance, \cite{Bouchaud2006,Cont2000})  of the form
\begin{equation}\label{pdfrpowtails}
\phi(r) \sim |r|^{-\mu-1} , \qquad r\to \pm \infty .
\end{equation}
Thus, we need in probability density function $\phi(r)$ of tick-by-tick returns $\{r_k\}$ \eqref{rkeqykhk}, possessing by the same power tail.
We suggest below that $\mu>\theta$ in order for the moments
\begin{equation}\label{r2notation}
\varepsilon_\theta := \text{E}\left[H_k^{-\theta}\right] < \infty
\end{equation}
of given order $\theta>0$ to remain finite.

One can prove rigorously that a sufficient condition for the probability density function
\begin{equation}
\phi(r) = \int_0^\infty \varphi(\eta) \psi(r\eta)\eta d\eta
\end{equation}
of the tick-by-tick returns $\{r_k\}$ \eqref{rkeqykhk} to possess a power law \eqref{pdfrpowtails} is
\begin{equation}
\varphi(\eta) = \eta^{\mu-1} \Phi(\eta) , \qquad \eta> 0 ,
\end{equation}
where $\Phi(\eta)$ is nonnegative continuous function, positive at $\eta=0$ ($\Phi(0)>0$), and decaying, at $\eta\to\infty$, faster than $\eta^{-\mu}$.
A convenient candidate for $\varphi(\eta)$ is
\begin{equation}
\varphi(\eta) = \frac{2 b^\mu}{2^{\mu/2}\Gamma(\mu/2)}~ \eta^{\mu-1} \exp\left(-\frac{b^2 \eta^2}{2}\right) .
\end{equation}
In this case, the probability density function $\phi(r)$ of the tick-by-tick returns is equal to
the Student distribution \eqref{pdfstudmu}. Correspondingly,
the moments $\varepsilon_\theta$ \eqref{r2notation} are given by the expression
\begin{equation}\label{barrthetaexpr}
\varepsilon_\theta = \frac{b^\theta \, \Gamma\left(\frac{\mu - \theta}{2}\right)}{2^{\theta/2} \, \sqrt{\pi}~ \Gamma\left(\frac{\mu}{2}\right)} .
\end{equation}

\subsection{Calculation of the correlation of absolute values of returns}

In order to calculate the autocorrelation function \eqref{absretcordef} of
the absolute values of returns, notice that
$|r_k|^\theta = |X_k|^\theta \, / H_k^\theta$, as it follows from \eqref{rkeqykhk}.
Accordingly, one has
\begin{equation}\label{exprkrkmtheta}
\begin{array}{c} \displaystyle
\text{E}\left[|r_k|^\theta \, |r_{k+m}|^\theta \right] =
\begin{cases}
\varepsilon_{2\theta} ~ \mathcal{F}(\theta,1) , & m = 0 ,
\\
\varepsilon^2_\theta ~ \mathcal{F}(\theta,\rho_m) , & m \geqslant 0 ,
\end{cases}
\\[6mm] \displaystyle
\text{E}^2\left[|r_k|^\theta\right] = \varepsilon^2_\theta  \mathcal{F}(\theta,0) ,
\end{array}
\end{equation}
where $\varepsilon_\theta$ is given by \eqref{barrthetaexpr}, while
\begin{equation}\label{rthdefs}
\begin{array}{c}\displaystyle
\mathcal{F}(\theta,\rho_m)  := \text{E}\left[|X_k|^\theta |X_{k+m}|^\theta\right] , \\[4mm] \displaystyle
\mathcal{F}(\theta,1) = \frac{2^\theta}{\sqrt{\pi}} \Gamma\left(\frac{1}{2} + \theta\right) , \qquad
\mathcal{F}(\theta,0) = \frac{2^\theta}{\pi} \Gamma^2\left(\frac{1+\theta}{2}\right) .
\end{array}
\end{equation}
After substitution \eqref{exprkrkmtheta} into \eqref{absretcordef}, we obtain
\begin{equation}\label{matcalfnorm}
\mathcal{A}_m(\theta) =
\begin{cases}
1 , & \rho = 1 ,
\\[4mm] \displaystyle
\frac{\varepsilon^2_\theta\,\widetilde{\mathcal{F}}(\theta,\rho_m)} {\varepsilon_{2\theta} \mathcal{F}(\theta,1) - \varepsilon^2_\theta \mathcal{F}(\theta,0)} , & \rho < 1 ,
\end{cases}
\end{equation}
where
\begin{equation}\label{ftildef}
\widetilde{\mathcal{F}}(\theta,\rho) : = \mathcal{F}(\theta,\rho)- \mathcal{F}(\theta,0) .
\end{equation}

To calculate the function $\mathcal{F}(\theta,\rho)$, we take
into account that the joint probability density function of
the Gaussian variables $X_k$ and $X_{k+m}$ is equal to
\begin{equation}
\psi(x_1,x_2;\rho_m) = \frac{1}{2\pi\sqrt{1-\rho_m^2}} \exp\left( - \frac{x_1^2 + x_2^2 - 2 \rho_m x_1 x_2}{2 (1-\rho_m^2)}\right) ,
\end{equation}
and we obtain
\begin{equation}\label{corabsthetxk}
\begin{array}{c}\displaystyle
\mathcal{F}(\theta,\rho) = \frac{\Gamma(1+\theta)}{\pi}  \left(1-\rho^2\right)^{\frac{1}{2}+\theta} \times
\\[4mm]\displaystyle
\int_0^{\pi/2} \left(\frac{1}{(1+\rho \sin u)^{1+\theta}} + \frac{1}{(1-\rho \sin u)^{1+\theta}}\right) \sin^\theta \! u \, du .
\end{array}
\end{equation}
In particular, for $\theta=1$, one has
\begin{equation}
\mathcal{F}(1,\rho) = \frac{2}{\pi} \left[ \sqrt{1-\rho^2} + \rho \arcsin \rho \right] .
\end{equation}

The correlation $\rho_m$ in its attainable range \eqref{rhomineq} is well approximated
by the first nonzero term of the Taylor expansion with respect to $\rho$
of the function $\widetilde{\mathcal{F}}(\theta,\rho)$ \eqref{ftildef}, \eqref{corabsthetxk},
which is given by
\begin{equation}\label{gthetrhosq}
\mathcal{G}(\theta,\rho) = g_\theta\, \rho^2 , \qquad  g_\theta : = \frac{\Gamma\left(\frac{1+\theta}{2}\right) \Gamma(\theta)}{\sqrt{\pi} \,\Gamma\left(\frac{\theta}{2}\right)} \theta^2 .
\end{equation}
The accuracy of the above quadratic approximation
is illustrated in figures~A3 and~A4. In particular, figure~A3 shows that
the quadratic approximation $\mathcal{G}(\theta,\rho)$ \eqref{gthetrhosq}
and the function $\widetilde{\mathcal{F}}(\theta,\rho)$ \eqref{ftildef} are
almost indistinguishable. Figure~A4 plots the ratio of these two functions in order to
provide a quantitive estimate of the relative errors. In the worst case for
$\theta=0.5$ and $\rho=0.5$, we observe a maximum 5\% relative error.

Replacing in \eqref{matcalfnorm} the function $\tilde{\mathcal{F}}(\theta,\rho_m)$ by its quadratic approximation \eqref{gthetrhosq}, we obtain
\begin{equation}\label{matcalfnormsquar}
\mathcal{A}(\theta,\rho) =
\begin{cases}
1 , & m = 0 ,
\\[4mm] \displaystyle
\frac{g_\theta \, \varepsilon^2_\theta} {\varepsilon_{2\theta}\, \mathcal{F}(\theta,1) - \varepsilon^2_\theta \, \mathcal{F}(\theta,0)}\, \rho^2 , & m\geqslant 1 ~.
\end{cases}
\end{equation}
Substituting in \eqref{matcalfnormsquar} the power law approximation \eqref{rhompower} of the autocorrelation function $\rho_m$ in place of the variable $\rho$, we obtain the power law \eqref{matcalfnormsquara}, where the factor $\chi$ is given by
\begin{equation}\label{betgam}
\chi := \digamma^2\left(\frac{\sigma}{2}\right) \frac{g_\theta \, \varepsilon^2_\theta} {\varepsilon_{2\theta} \, \mathcal{F}(\theta,1) - \varepsilon^2_\theta \,\mathcal{F}(\theta,0)} .
\end{equation}

\pagebreak

\section{Statistical description of instantaneous returns}

\setcounter{equation}{0}
\setcounter{theorem}{0}
\setcounter{remark}{0}
\setcounter{lemma}{0}
\setcounter{definition}{0}
\setcounter{example}{0}
\setcounter{supposition}{0}

\subsection{Correlation function of instantaneous returns}

We outline in this subsection one of the possible derivations of relation \eqref{btaugenexpr} for
the continuous part $B(\tau)$ of the correlation function $K(\tau)$ \eqref{natcovthruaucorm}
of the instantaneous returns $R(t)$ \eqref{returnsdef}.

Notice, first of all, that
\begin{equation}\label{btauthruk}
\nu \varepsilon_2 \, B(\tau) \equiv K(\tau) \qquad \forall \tau\neq 0 .
\end{equation}
Using definition \eqref{returnsdef} together with the ordering rule \eqref{tkordering}
of the trade instants $\{t_k\}$, we rewrite relation \eqref{btauthruk} in the form
\begin{equation}
K(\tau) = \varepsilon_2 \sum_{m=1}^\infty \mathcal{B}_m \, \text{E}\left[ \sum_k \delta(t-t_k) \delta(t-t_{k+m}+\tau)\right] , \qquad \forall \tau> 0 ,
\label{heyjuki}
\end{equation}
where $\{\mathcal{B}_m\}$ are the  correlations \eqref{rkcorfuncall} of the tick-by-tick returns.
We keep in mind that, within the assumptions of our microstructure return model,
the sequences $\{r_k\}$ and $\{t_k\}$ are mutually statistically independent.
In view of identity \eqref{btauthruk}, we rewrite (\ref{heyjuki}) in the form
\begin{equation}\label{ktaumoreo}
\nu B(\tau) \equiv  \sum_{m=1}^\infty \mathcal{B}_m \, \text{E}\left[ \sum_k \delta(t-t_k) \delta(t-t_{k+m}+\tau)\right] , \qquad \forall \tau> 0 .
\end{equation}

Applying the linear operator
\begin{equation}
\hat{\mathcal{L}} := \frac{1}{T} \int_0^T \dots dt
\end{equation}
to both sides of equality \eqref{ktaumoreo}
and using the commutativity between linear operations and statistical expectations, we obtain
\begin{equation}
\nu B(\tau) \equiv \sum_{m=1}^\infty \mathcal{B}_m \, \text{E}\left[ \frac{1}{T} \int_0^T \sum_{k=1}^{N(T)} \delta(t-t_k) \delta(t-t_{k+m}+\tau) dt \right] , \qquad \forall \tau> 0 ,
\end{equation}
where $N(T)$ is the random number of instants $\{t_k\}$ belonging to
the time interval $t\in(0,T)$. After integration, one has
\begin{equation}
\nu B(\tau) \equiv  \sum_{m=1}^\infty \mathcal{B}_m \, \text{E}\left[ \frac{1}{T} \sum_{k=1}^{N(T)} \delta(t_k-t_{k+m}+\tau) dt \right] , \qquad \forall \tau> 0~ .
\end{equation}

Consider the limit of the last relation for $T\to \infty$. Using the Law of Large Numbers,
one may replace, at $T\to\infty$, the random number $N(T)$ by its expected value
$\text{E}\left[N(T)\right] = \nu T$. This yields
\begin{equation}
\nu B(\tau) \equiv  \sum_{m=1}^\infty \mathcal{B}_m \,\lim_{T\to\infty} \frac{\nu T}{T} \text{E}\left[\delta(\tau - \tau_{k,m}) \right] , \qquad \forall \tau> 0 ,
\label{wrhj5iko}
\end{equation}
where we have used notation \eqref{taukmdef}. Using the fact
that the expectation in (\ref{wrhj5iko}) gives by definition
the probability density function of the random interval $\tau_{k,m}$,
$f_m(\tau) := \text{E}\left[\delta(\tau - \tau_{k,m}) \right]$,
we finally obtain
\begin{equation}
B(\tau) \equiv \sum_{m=1}^\infty B_m  f_m(\tau) \qquad \forall \tau> 0~ .
\end{equation}
Taking into account the evenness of the correlation functions $B_m$'s,
we obtain relation \eqref{btaugenexpr}.

\subsection{Laplace images of some generalized gamma distributions}

We present analytical expressions of the Laplace images
of generalized gamma distributions.

In the case $\beta=1/2$, one has
\begin{equation}\label{hatonttwoexpl}
\hat{f}(s) = \hat{g}_{1/2}\left(\frac{s}{\lambda_{1/2}(\vartheta)};\vartheta\right) , \qquad \lambda_{1/2}(\vartheta ) = 2\vartheta(1+2\vartheta) ,
\end{equation}
and
\begin{equation}\label{gengamonetwolap}
\hat{g}_{1/2}(s;\vartheta) = \frac{1}{2 \Gamma(2\vartheta) s^\vartheta} \sum_{k=0}^1 \frac{(-1)^k}{s^\frac{k}{2}}\Gamma\left(\vartheta+\frac{k}{2}\right) \Phi\left(\frac{k}{2}+\vartheta, \frac{2 k+1}{2}, \frac{1}{4 s} \right) .
\end{equation}
Here $\Phi(a,b,z)$ is the Kummer function (confluent hypergeometric function).

In the case $\beta=1/3$, one has
\begin{equation}\label{hatonthreeexpl}
\hat{f}(s) = \hat{g}_{1/3}\left(\frac{s}{\lambda_{1/3}(\vartheta)};\vartheta\right) , \qquad \lambda_{1/3}(\vartheta ) = 3\vartheta(1+3\vartheta) (2+3\vartheta) ,
\end{equation}
and
\begin{equation}\label{gengamonethreelap}
\begin{array}{c}\displaystyle
\hat{g}_{1/3}(s;\vartheta) = \frac{1}{6 \Gamma(3\vartheta) s^\vartheta}  \sum_{k=0}^2 \frac{(-1)^k}{s^{\frac{k}{3}}} 2^\frac{(k+1)(2-k)}{2} \Gamma\left(\vartheta + \frac{k}{3}\right) \times
\\[6mm]\displaystyle
\mathstrut_1\! H_2 \left(\vartheta + \frac{k}{3}; \frac{(2k+1)^2+7}{24}, \frac{41-(2k-5)^2}{24}; -\frac{1}{27 s} \right) .
\end{array}
\end{equation}
Here $\mathstrut_P\! H_Q(\boldsymbol{a};\boldsymbol{b};z)$ is the generalized hypergeometric series, while $\boldsymbol{a}$ and $\boldsymbol{b}$ are, correspondingly, vectors of lengths $P$ and $Q$.

An analogous analytical expression for the Laplace image $\hat{f}(s)$
of the probability density function $f(\tau)$ taking the form of
the generalized gamma distribution with $\beta=2/3$ is given by
\begin{equation}\label{thethreelam}
\hat{f}(s) = \hat{g}_{2/3}\left(\frac{s}{\lambda_{2/3}(\vartheta)};\vartheta\right) , \qquad \lambda_{2/3}(\vartheta ) = \frac{\Gamma(3 (\vartheta+1)/2)} {\Gamma(3 \vartheta/2)} ,
\end{equation}
and
\begin{equation}\label{twothree}
\begin{array}{c}\displaystyle
\hat{g}_{2/3}(s;\vartheta) = \frac{1}{3 \Gamma(3\vartheta/2) s^\vartheta}  \sum_{k=0}^2 \frac{(-1)^k}{s^{\frac{2k}{3}}} 2^\frac{(k+1)(2-k)}{2} \Gamma\left(\vartheta + \frac{2k}{3}\right) \times
\\[6mm]\displaystyle
\mathstrut_2 H_2 \left(\frac{\vartheta}{2}+a_1(k), \frac{\vartheta}{2} +a_2(k); \frac{(2k+1)^2+7}{24}, \frac{41-(2k-5)^2}{24}; -\frac{4}{27 s^2} \right) ,
\\[6mm]\displaystyle
a_1(k) :=  \frac{(6k-5)^2+47}{144} , \qquad  a_2(k) := \frac{k(13-3k)}{12} .
\end{array}
\end{equation}

\subsection{Spectral representations of the expectations of quadratic returns}

In the numerical calculations of the correlation functions of returns,
we use relations that express the correlation functions through the corresponding power spectra. For instance, we calculate the continuous part $B(\tau)$ of the correlation function $K(\tau)$
of the instantaneous returns $R(t)$ by using
the inverse Fourier transform
\begin{equation}
B(\tau) = \frac{1}{\pi} \int_0^\infty \tilde{B}(\omega) \cos(\omega\tau) d\omega .
\end{equation}
Analogously, using the fact that the spectrum $\tilde{K}_\Delta(\omega)$ of the correlation function $K_\Delta(\tau)$ \eqref{adelcorconv} is given by
\begin{equation}\label{kdelmtdeldef}
\begin{array}{c} \displaystyle
\tilde{K}_\Delta(\omega) = \varepsilon_2 \,\tilde{A}(\omega) \, \tilde{T}_\Delta(\omega)  = \frac{\varepsilon_2}{2\pi} \tilde{\mathcal{T}}_\Delta(\omega) \left[1 + \tilde{B}(\omega)\right] ,
\\[4mm] \displaystyle
\tilde{\mathcal{T}}_\Delta(\omega) := 2 \int_0^\infty \mathcal{T}_\Delta(\tau)  \cos(\omega \tau) d\tau = \frac{4}{\omega^2} \sin^2 \left(\frac{\omega\Delta}{2}\right)  ,
\end{array}
\end{equation}
we obtain the following relation
\begin{equation}\label{adelexpr}
K_\Delta(\tau) = \frac{\varepsilon_2 }{\pi} \int_0^\infty \tilde{\mathcal{T}}_\Delta(\omega) \left[1+ \tilde{B}(\omega) \right] \cos(\omega \tau) d\omega ~,
\end{equation}
which is convenient for numerical calculations.
Similarly, we have used in our numerical calculations the relation
\begin{equation}\label{donetwospectr}
D_{1,2}(\Delta) = \frac{1}{\pi\Delta} \int_0^\infty \left[1+\tilde{B}(\omega)\right] \tilde{\kappa}(\omega) \tilde{\mathcal{T}}_\Delta(\omega) d\omega ,
\end{equation}
which is equivalent to relation \eqref{d12timeint}, where $\tilde{\mathcal{T}}_\Delta(\omega)$ is defined in \eqref{kdelmtdeldef}, while
\begin{equation}
\tilde{\kappa}(\omega) = 2 \int_0^\infty \kappa(\tau) \cos(\omega\tau) d\tau = \exp\left( - \frac{\omega^2 \lambda^2}{2} \right)~ .
\end{equation}

\clearpage

\begin{quote}
\centerline{
\includegraphics[width=14cm]{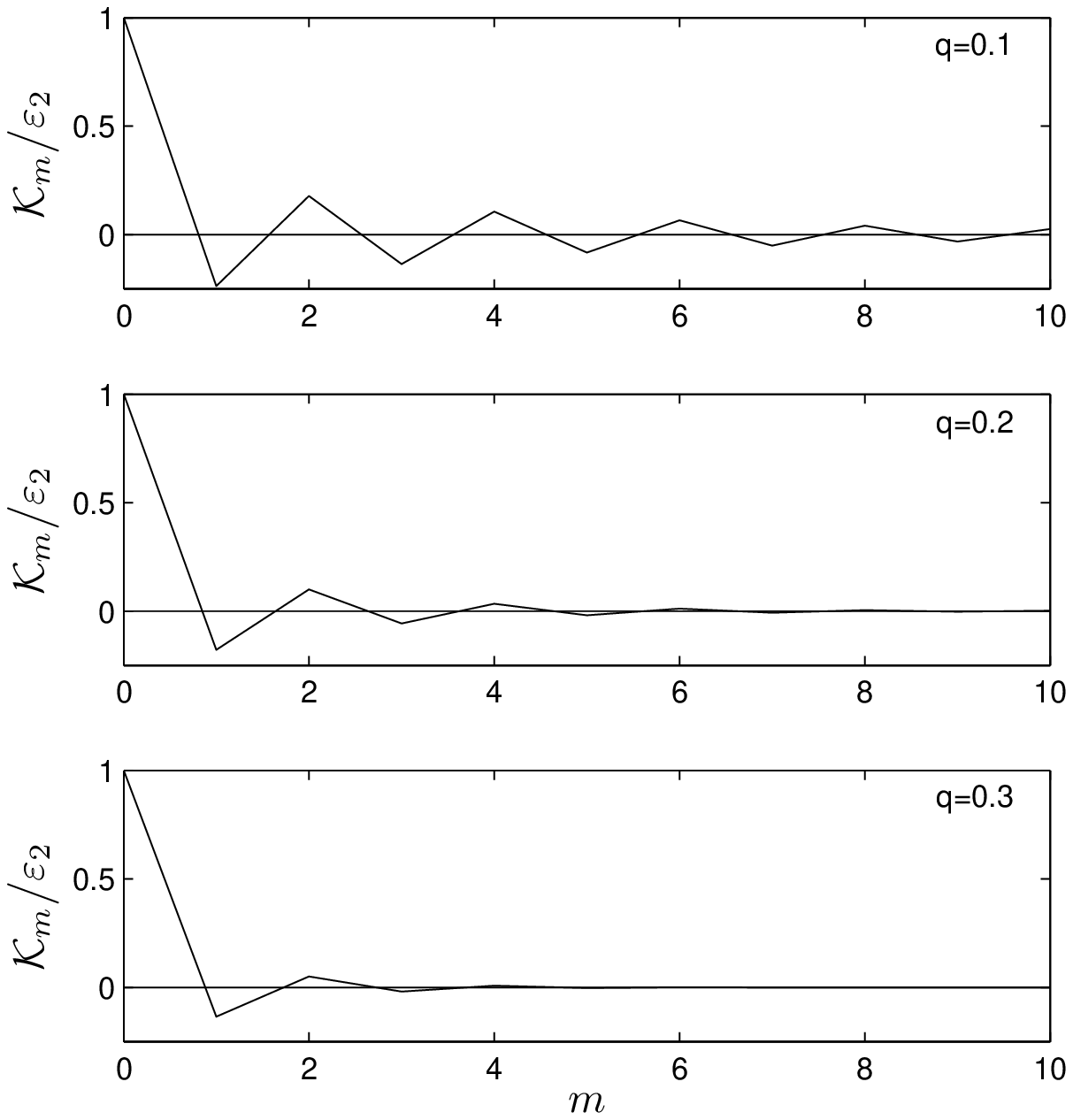}}
{\bf Fig.~1:} \small{Plots of the tick-by-tick returns correlation function $\mathcal{K}_m$, for $\alpha=0.1$, $\mu=4$ and $q=0.1; 0.2; 0.3$ (top to bottom).}
\end{quote}

\clearpage

\begin{quote}
\centerline{
\includegraphics[width=13cm]{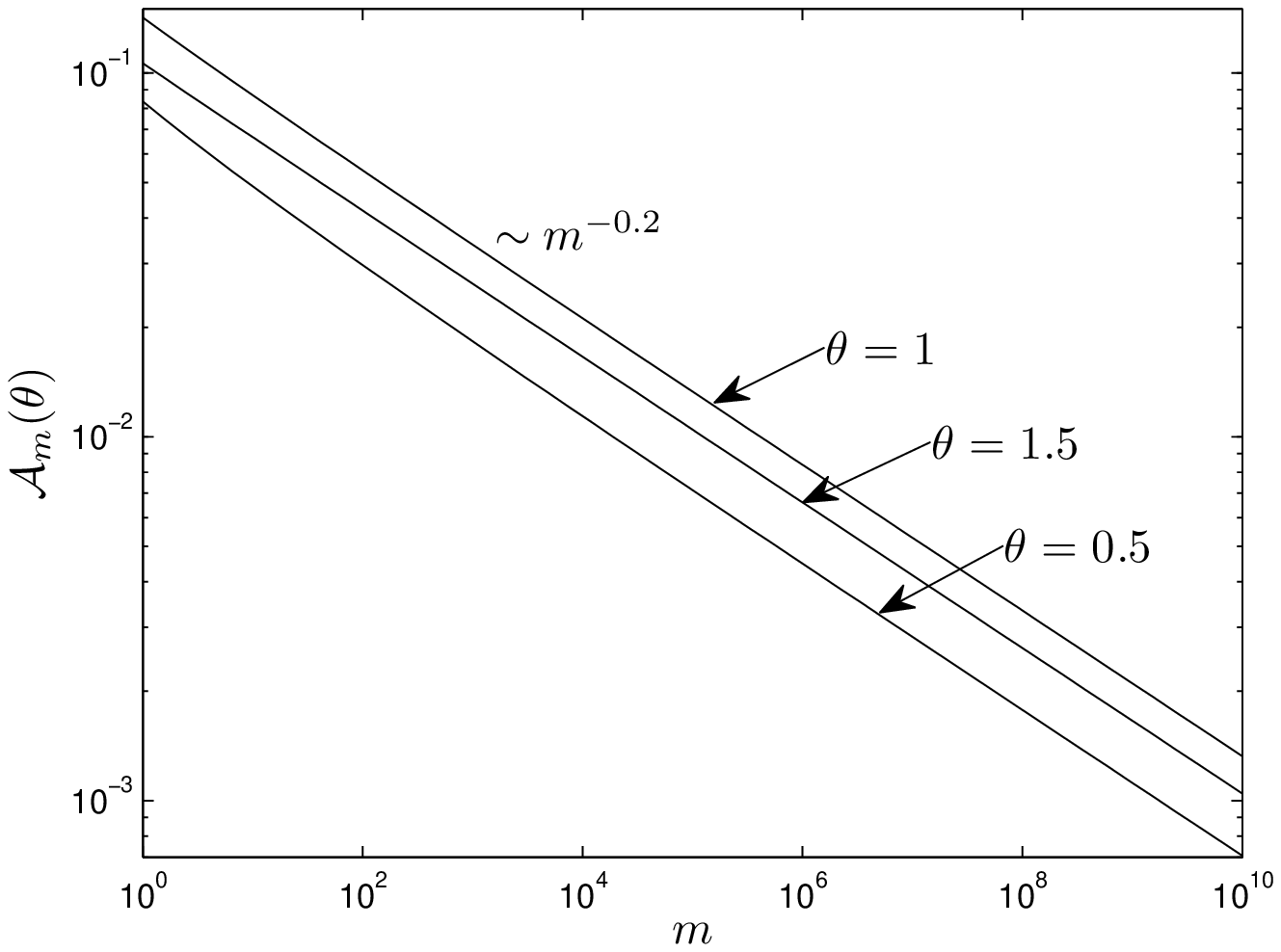}}
{\bf Fig.~2:} \small{Exact dependence of the autocorrelation function $\mathcal{A}_m(\theta)$ \eqref{absretcordef}, for $\mu=4$, $\sigma=0.2$ and $\theta=0.5; 1; 1.5$, illustrating the high accuracy of the approximate power law \eqref{matcalfnormsquara} qualified as a straight line in this log-log plot.}
\end{quote}

\clearpage

\begin{quote}
\centerline{
\includegraphics[width=15cm]{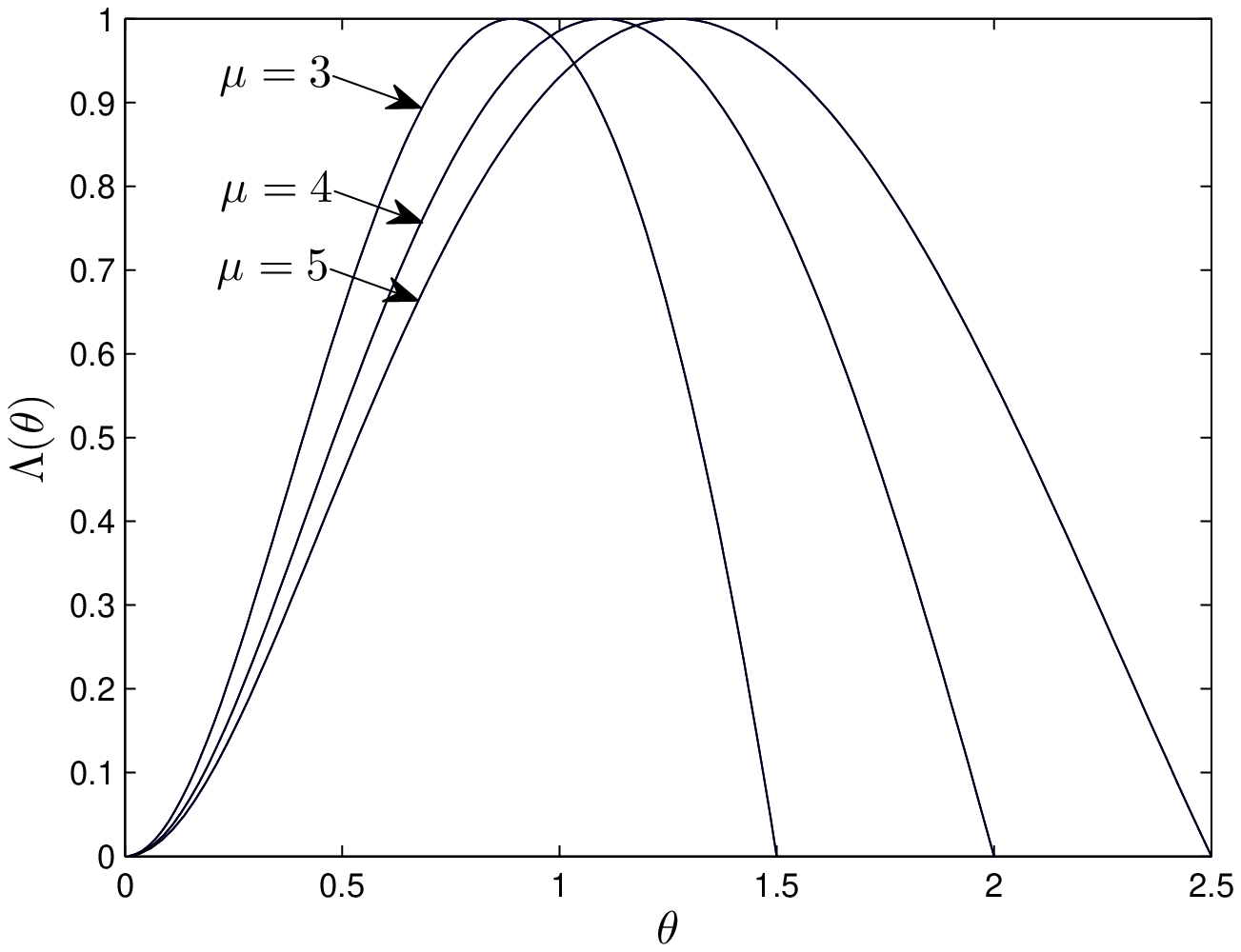}}
{\bf Fig.~3:} \small{Dependence as a function of $\theta$
of the normalized correlation function \eqref{mathdmaxratio} for different values
of the exponent $\mu$ of the Student distribution $\phi(r)$ \eqref{pdfstudmu}.}
\end{quote}

\clearpage

\begin{quote}
\centerline{
\includegraphics[width=15cm]{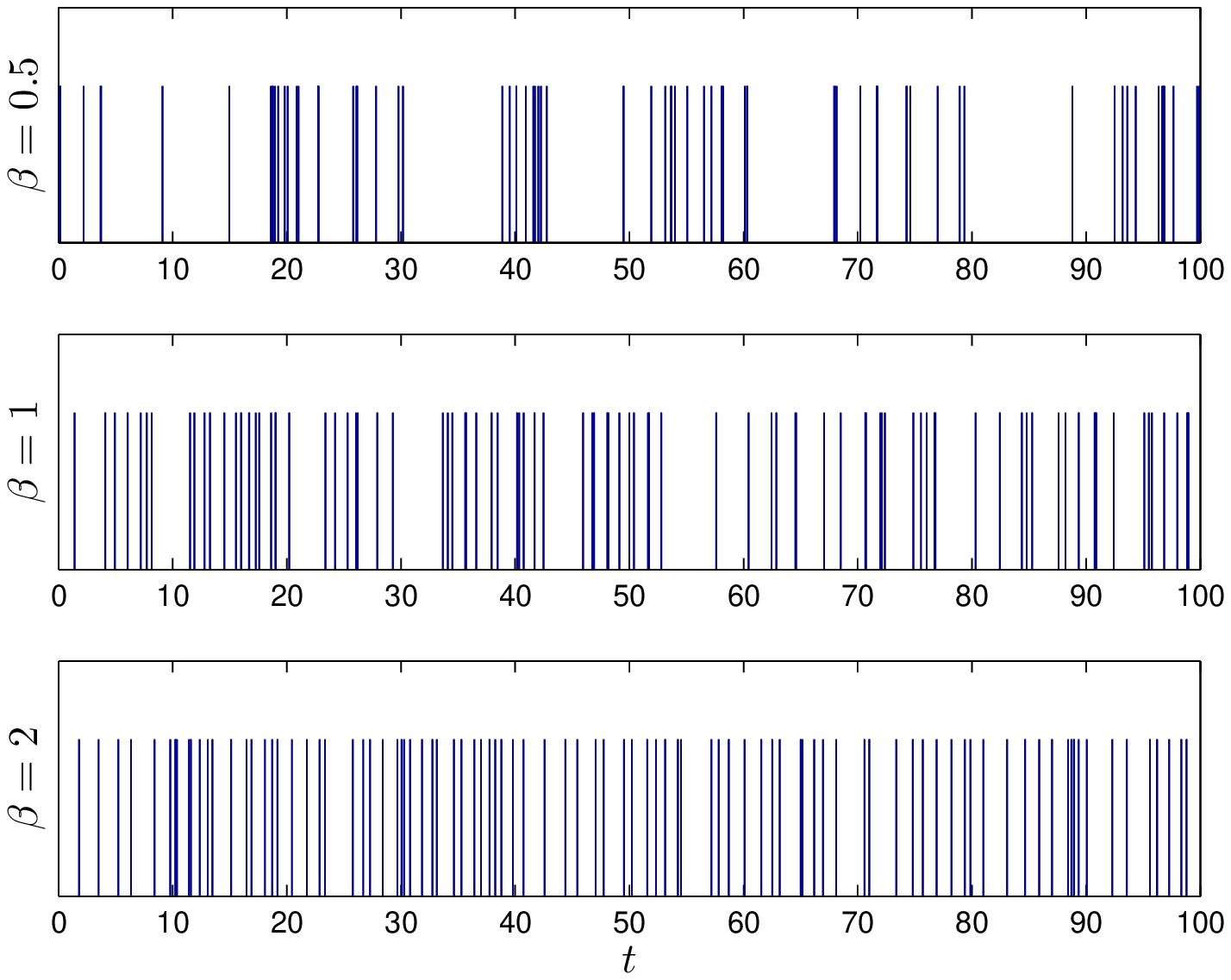}}
{\bf Fig.~4:} \small{Representation along the time axis of the time sequences $\{t_k\}$
of trades with $t_k = \sum_{i=1}^k \tau_i$,  where the inter-trade intervals $\tau_i$'s are drawn from the
Weibull distribution with shape parameters $\beta=0.5; 1; 2$ (top to bottom).  The corresponding values of the scaling parameter $\lambda_w(\beta)$ are given by expression \eqref{meannormcond}.
The value $\beta=1$ corresponds to pure Poisson memoryless sequences $\{t_k\}$.
For $\beta=0.5$, one can observe some events clustering. In contrast, for $\beta=2$,
the time sequence is more regular, akin to quasi-periodic.}
\end{quote}

\clearpage

\begin{quote}
\centerline{
\includegraphics[width=13cm]{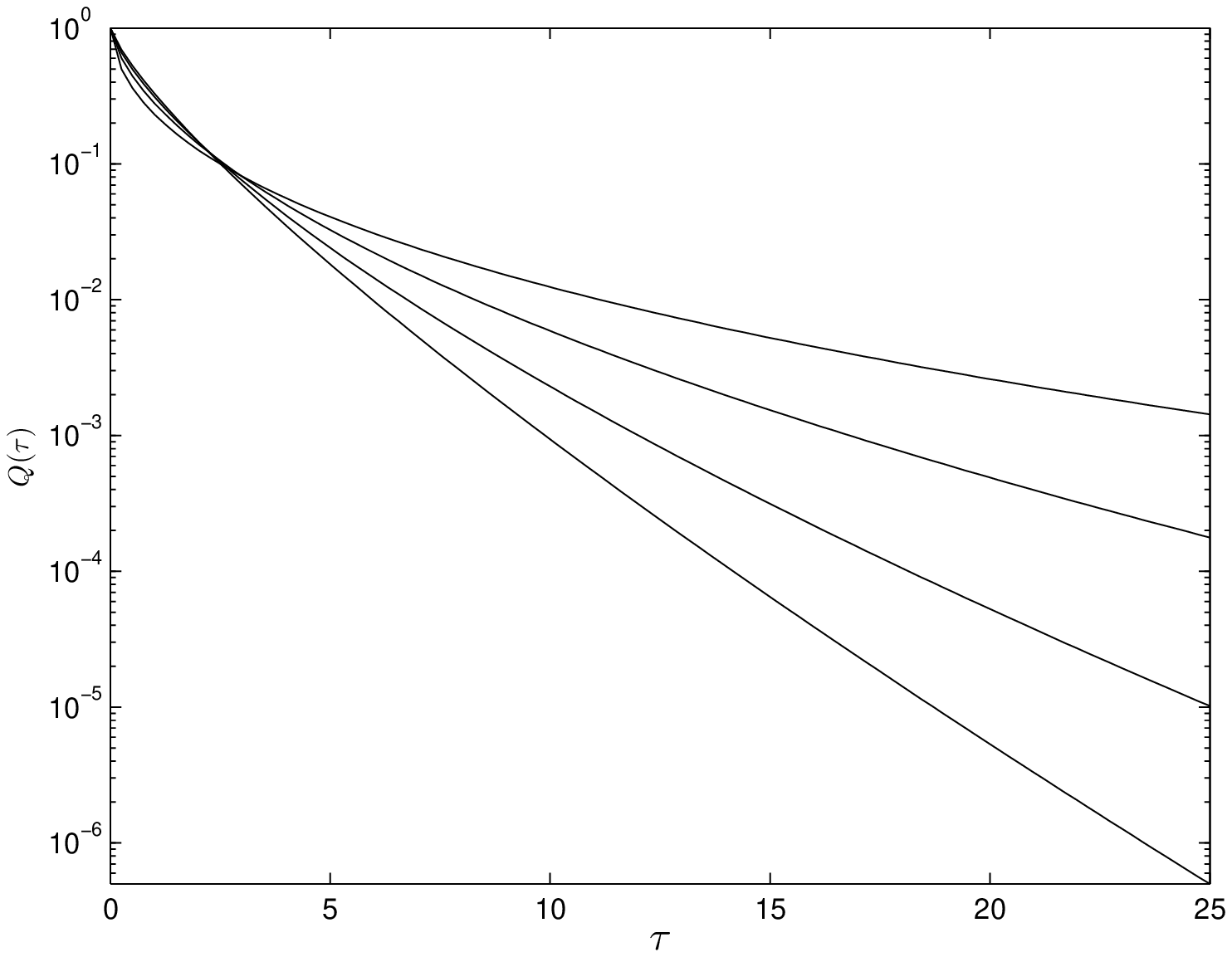}}
{\bf Fig.~5:} \small{Survival functions of the Weibull distribution \eqref{cweibdef} for $\beta=0.8$
(lower curve)  and of the generalized gamma distribution \eqref{weibsurvgendef}, for the same $\vartheta=0.8$ and for $\beta=1/3$, $1/2$ and $\beta=2/3$ (first, second and third curves from top at the right).}
\end{quote}

\clearpage

\begin{quote}
\centerline{
\includegraphics[width=13cm]{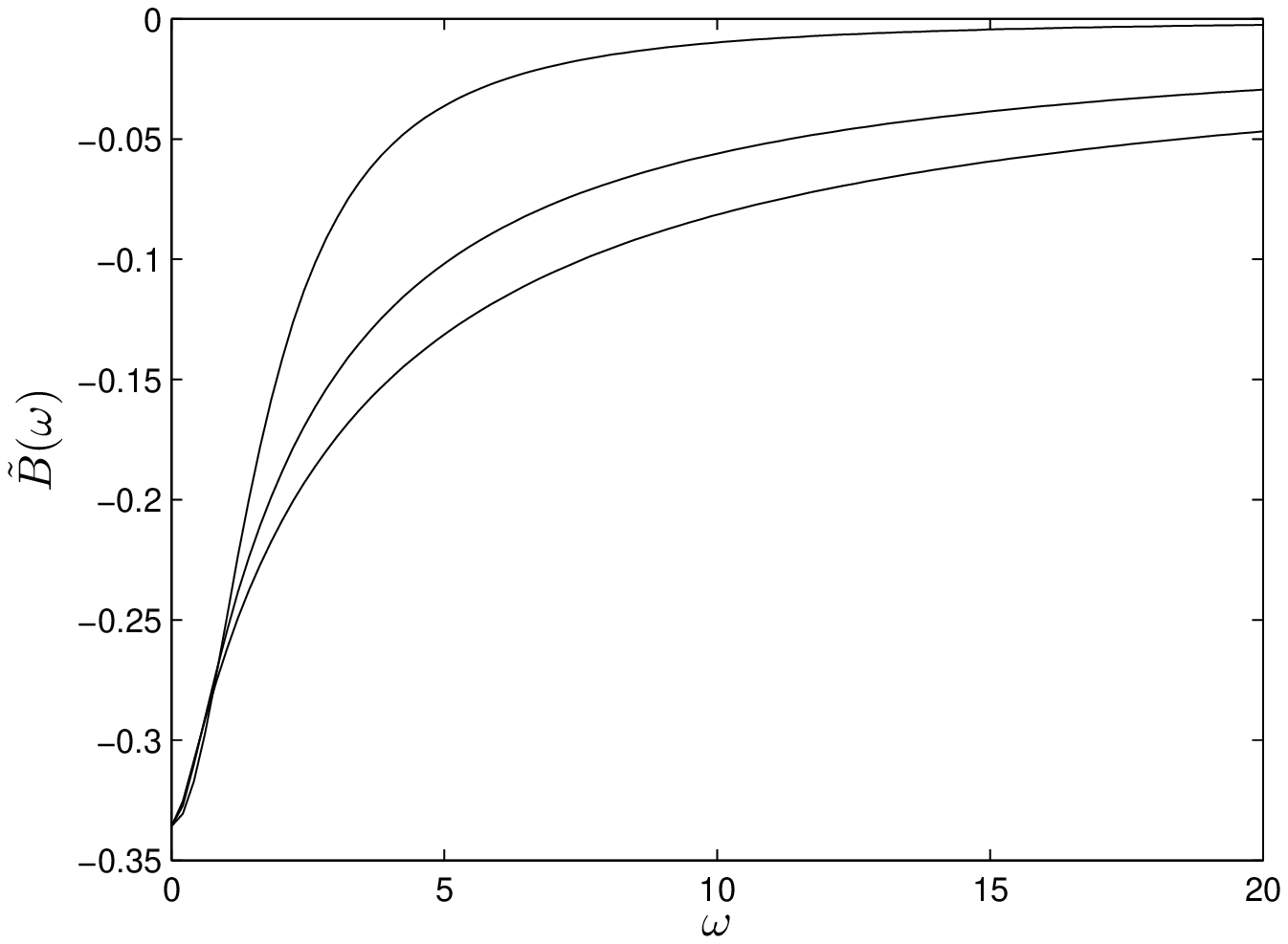}}
{\bf Fig.~6:} \small{Comparison of the function $\tilde{B}(\omega)$ \eqref{somintrepr}
(top curve)
obtained for a Poissonian statistics of the tick-by-tick times $\{t_k\}$ (\ref{pdflapois})
to the two functions $\tilde{B}(\omega)$ for the GGDs for $\beta=1/2; 2/3$
(first and second curves from bottom), for the same $\vartheta=0.8$.
The other parameters are $\alpha=0.1$, $\mu=4$ and $q=0.1$.}
\end{quote}

\clearpage

\begin{quote}
\centerline{
\includegraphics[width=13cm]{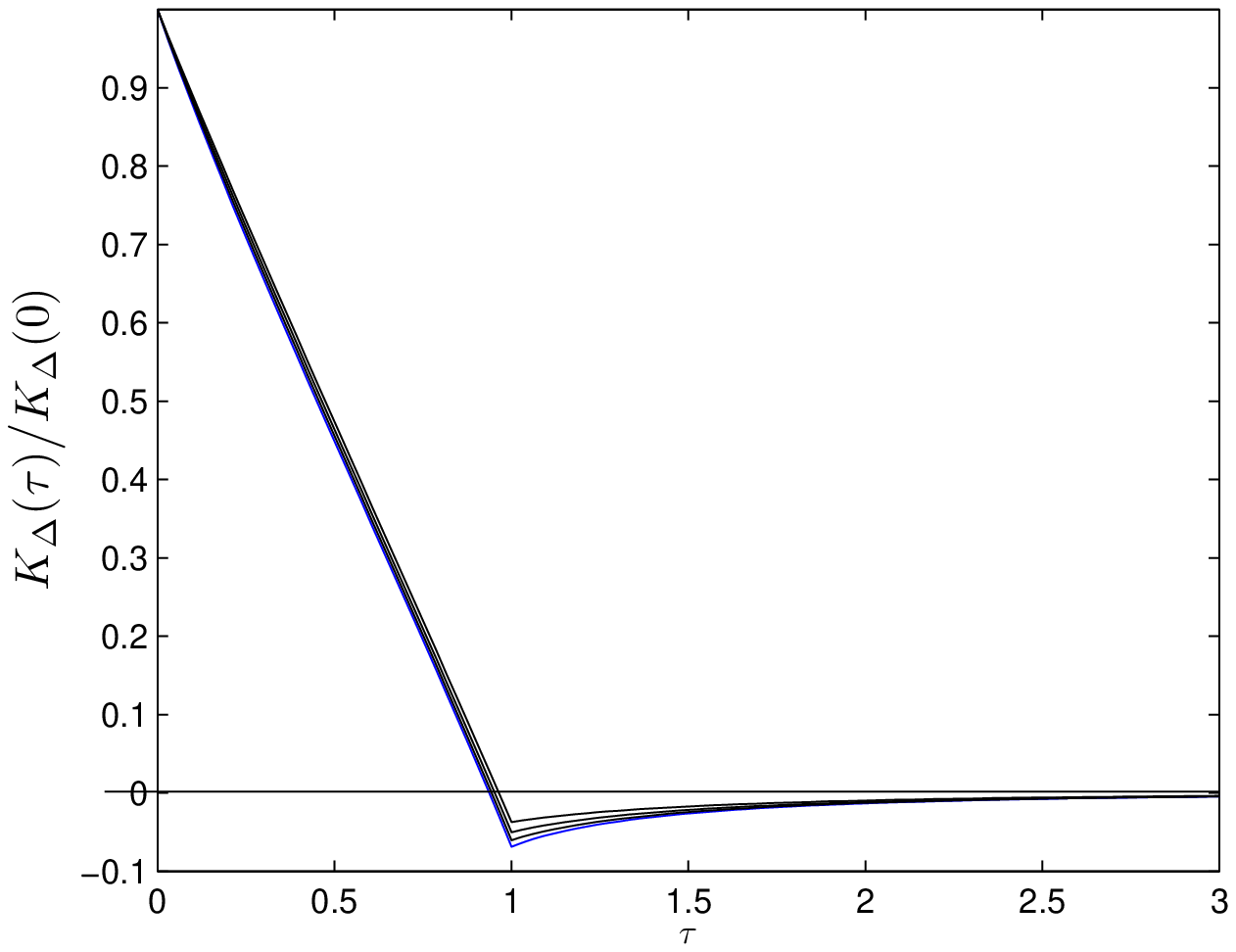}}
{\bf Fig.~7:} \small{Normalized correlation function $K_\Delta(\tau)/K_\Delta(0)$
of the $\Delta$-scale returns $R_\Delta(t)$ in calendar time, for $\Delta=1$ and $q=0; 0.1; 0.2; 0.3$ (bottom to top). The other parameters are $\alpha=0.1$, $\mu=4$,  $\vartheta=0.8$ and $\beta = 2/3$. It is clearly seen that, regardless of the value of the bounce distortion probability $q$ ($q<0.5$), the returns $R_\Delta(t)$ in calendar time are negatively short-ranged correlated for $\tau >1$. }
\end{quote}

\clearpage

\begin{quote}
\centerline{
\includegraphics[width=16cm]{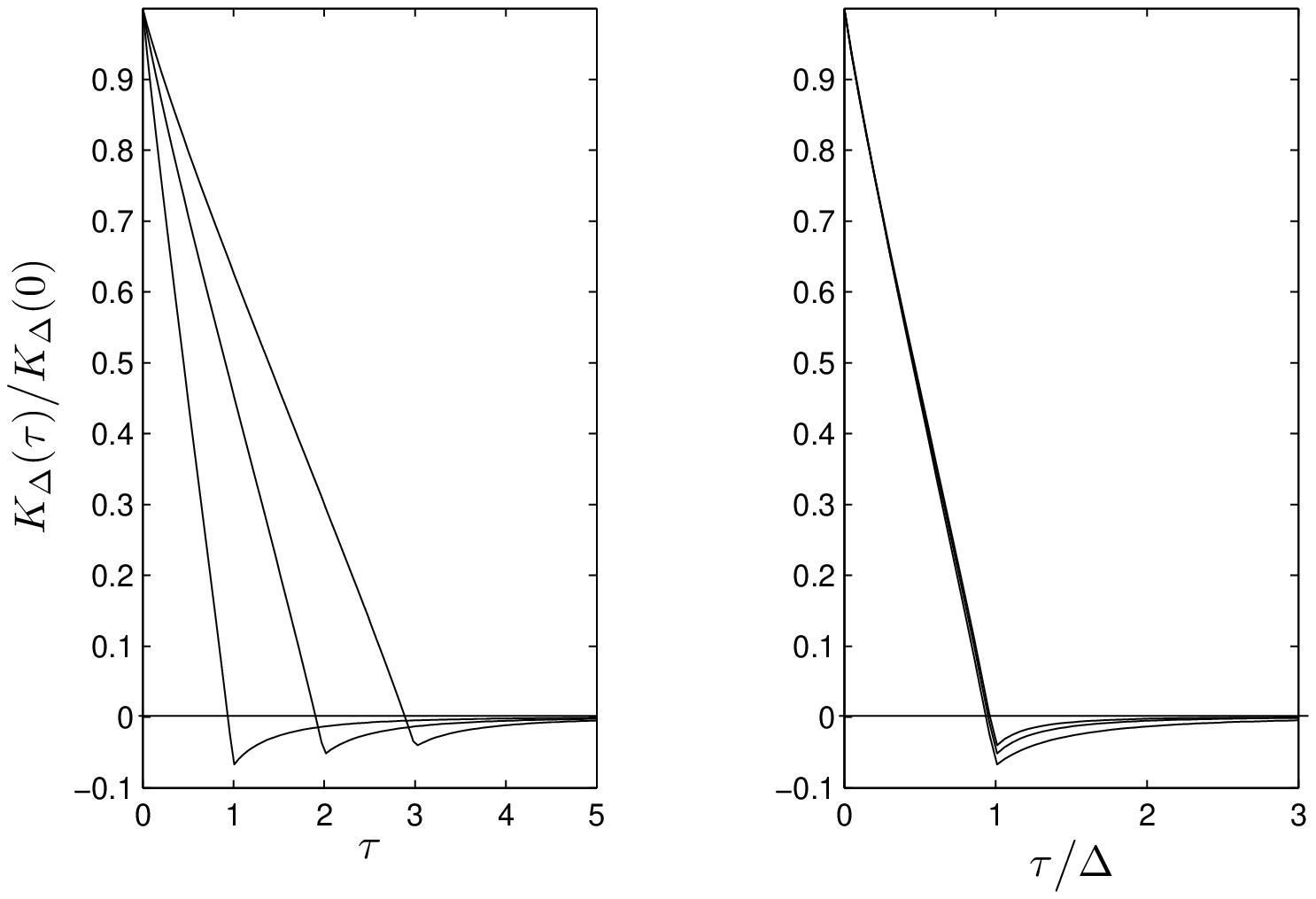}}
{\bf Fig.~8:} \small{Same as figure~7 for different values of $\Delta$.
Left panel: $\Delta=1; 2; 3$ (left to right) and $q=0$. Right panel: $\Delta=1; 2; 3$ and $q=0$, depicted in the calendar time scaled to $\Delta$.}
\end{quote}

\clearpage

\begin{quote}
\centerline{
\includegraphics[width=13cm]{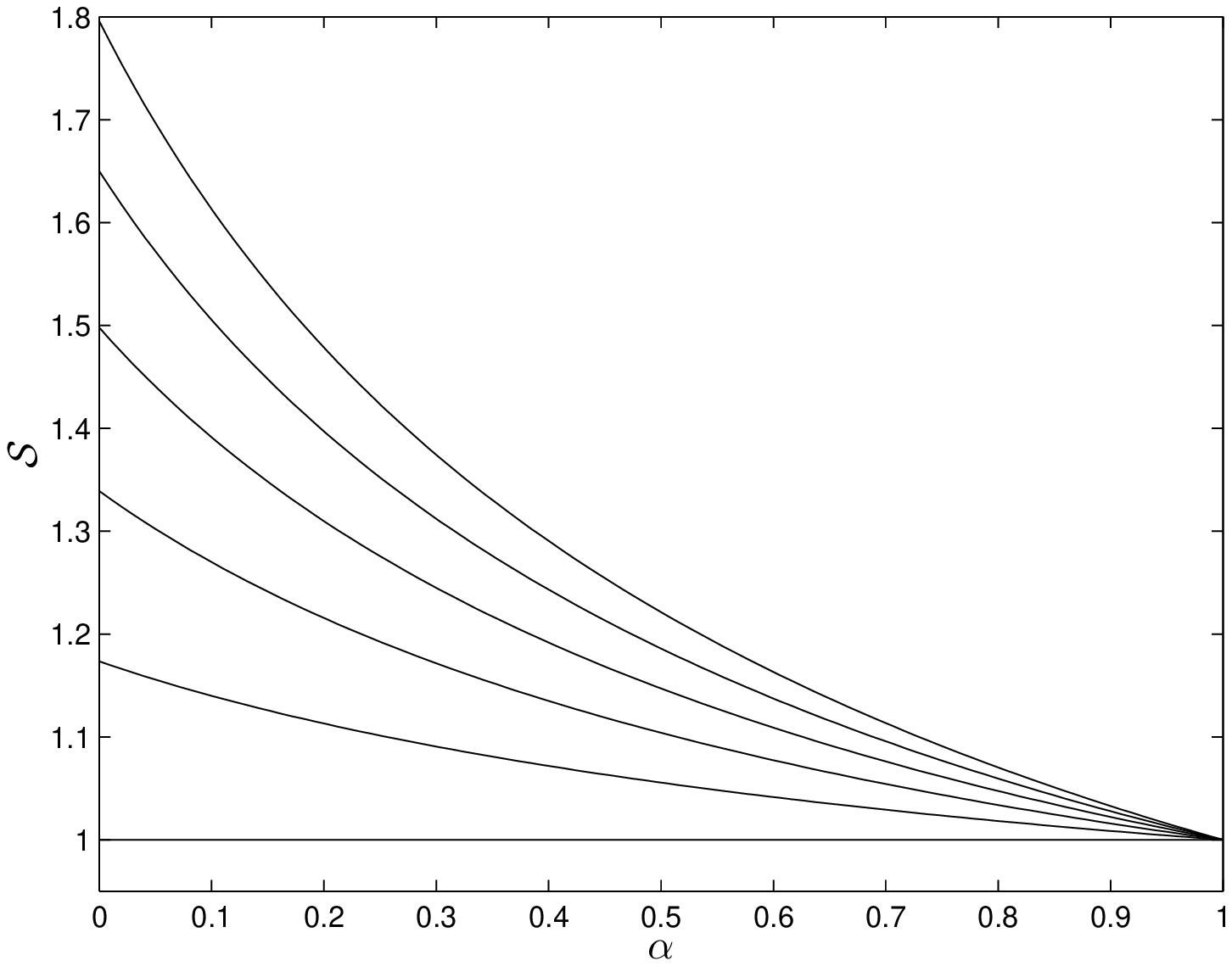}}
{\bf Fig.~9:} \small{Dependence of the microstructure noise strength $\mathcal{S}$ as a function of
the exponent $\alpha$, the exponent of the power law correlation \eqref{rhompower} of
the auxiliary ARFIMA process $\{X_k\}$. Here, $\mu=4$. Top to bottom: $q=0; 0.1; 0.2; 0.3; 0.4; 0.5$}
\end{quote}

\clearpage

\begin{quote}
\centerline{
\includegraphics[width=13cm]{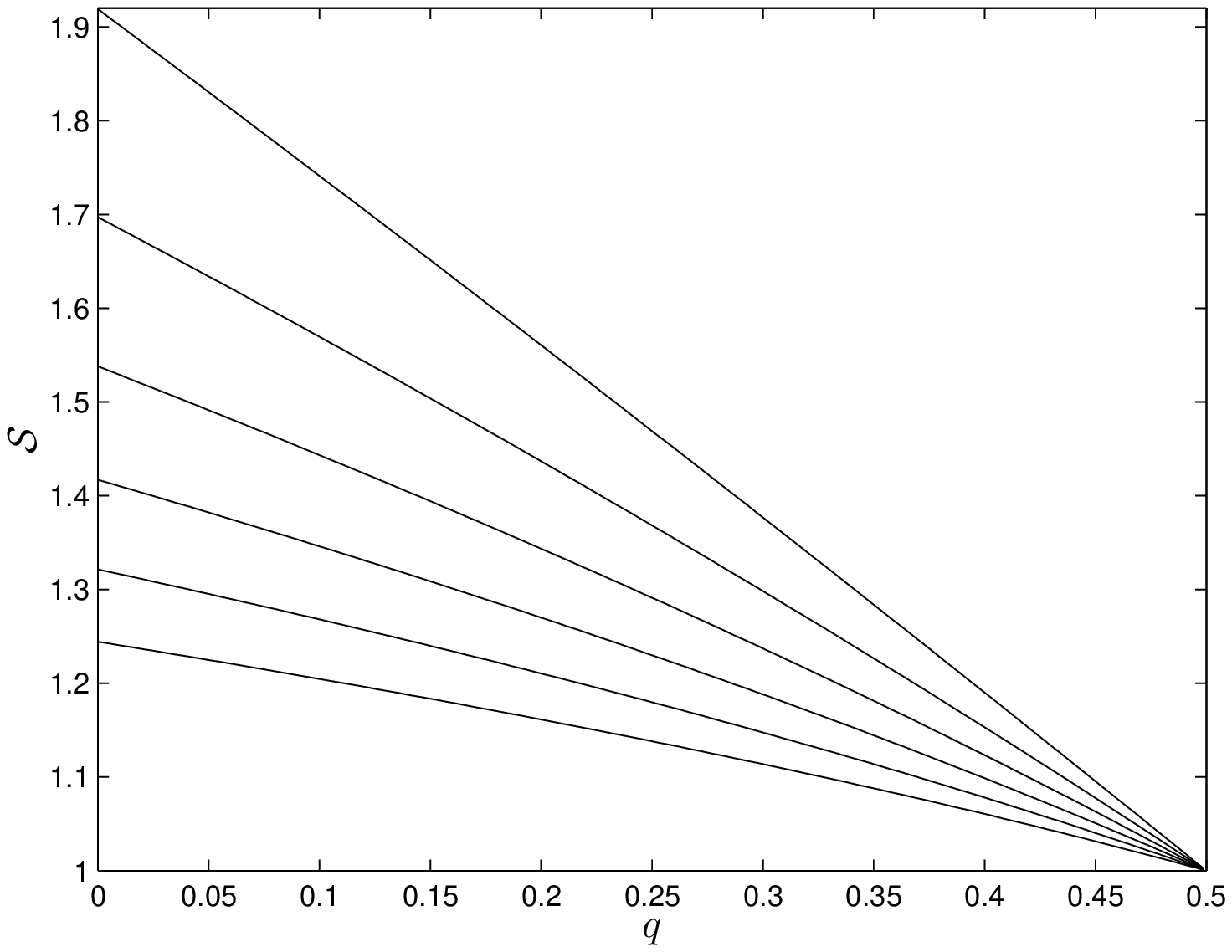}}
{\bf Fig.~10:} \small{Dependence of the microstructure noise strength $\mathcal{S}$ as a function of
the bounce distortion probability $q$, for $\mu=5$. Top to bottom: $\alpha=0; 0.1; 0.2; 0.3; 0.4; 0.5$}
\end{quote}

\clearpage

\begin{quote}
\centerline{
\includegraphics[width=13cm]{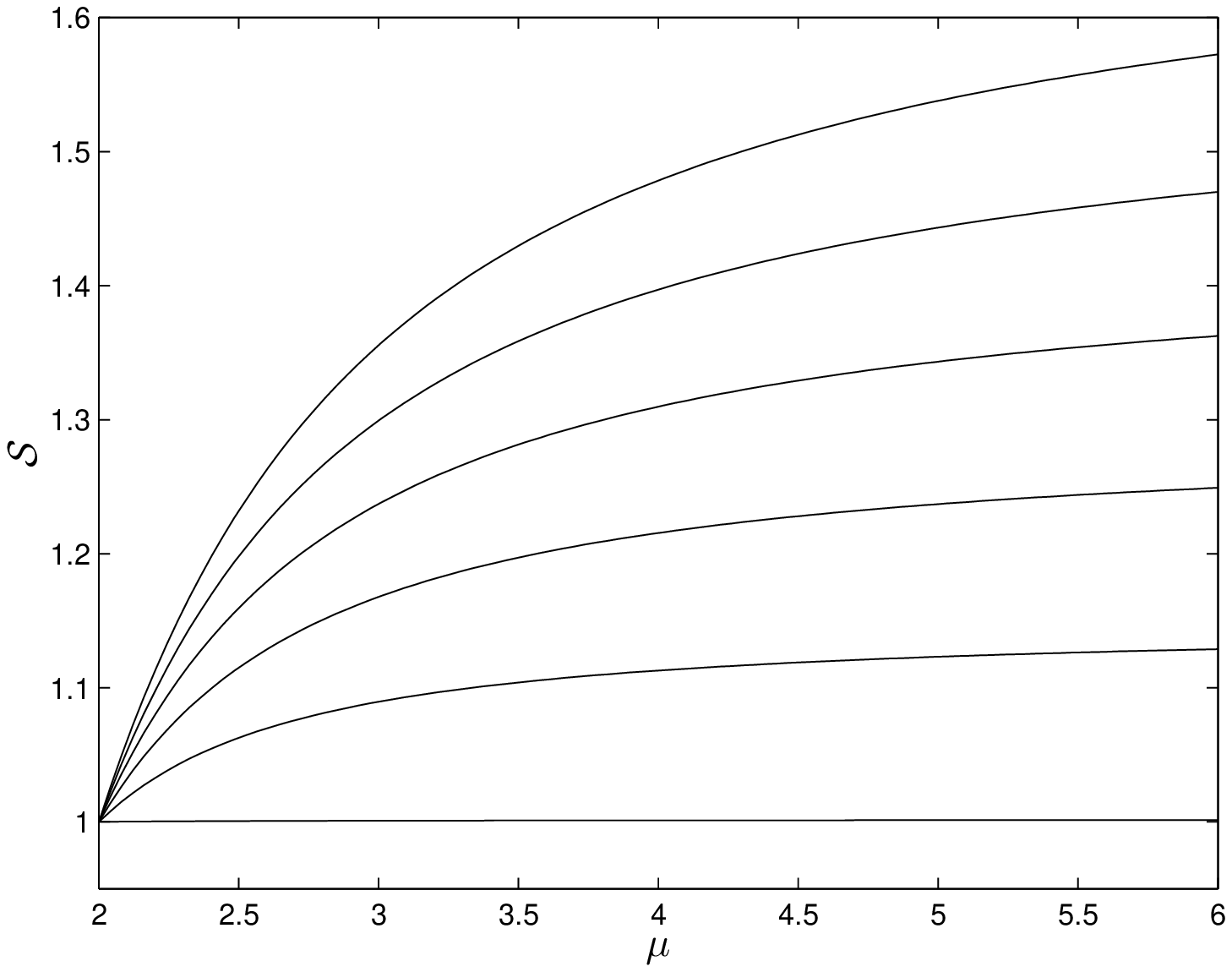}}
{\bf Fig.~11:} \small{Dependence of the microstructure noise strength $\mathcal{S}$ as a function of
the exponent $\mu$, the exponent of the power law describing the tail of
the probability density function $\phi(r)$ \eqref{pdfstudmu} of the tick-by-tick returns.
Here, $\alpha=0.2$. Top to bottom: $q=0; 0.1; 0.2; 0.3; 0.4; 0.5$}
\end{quote}

\clearpage

\begin{quote}
\centerline{
\includegraphics[width=13cm]{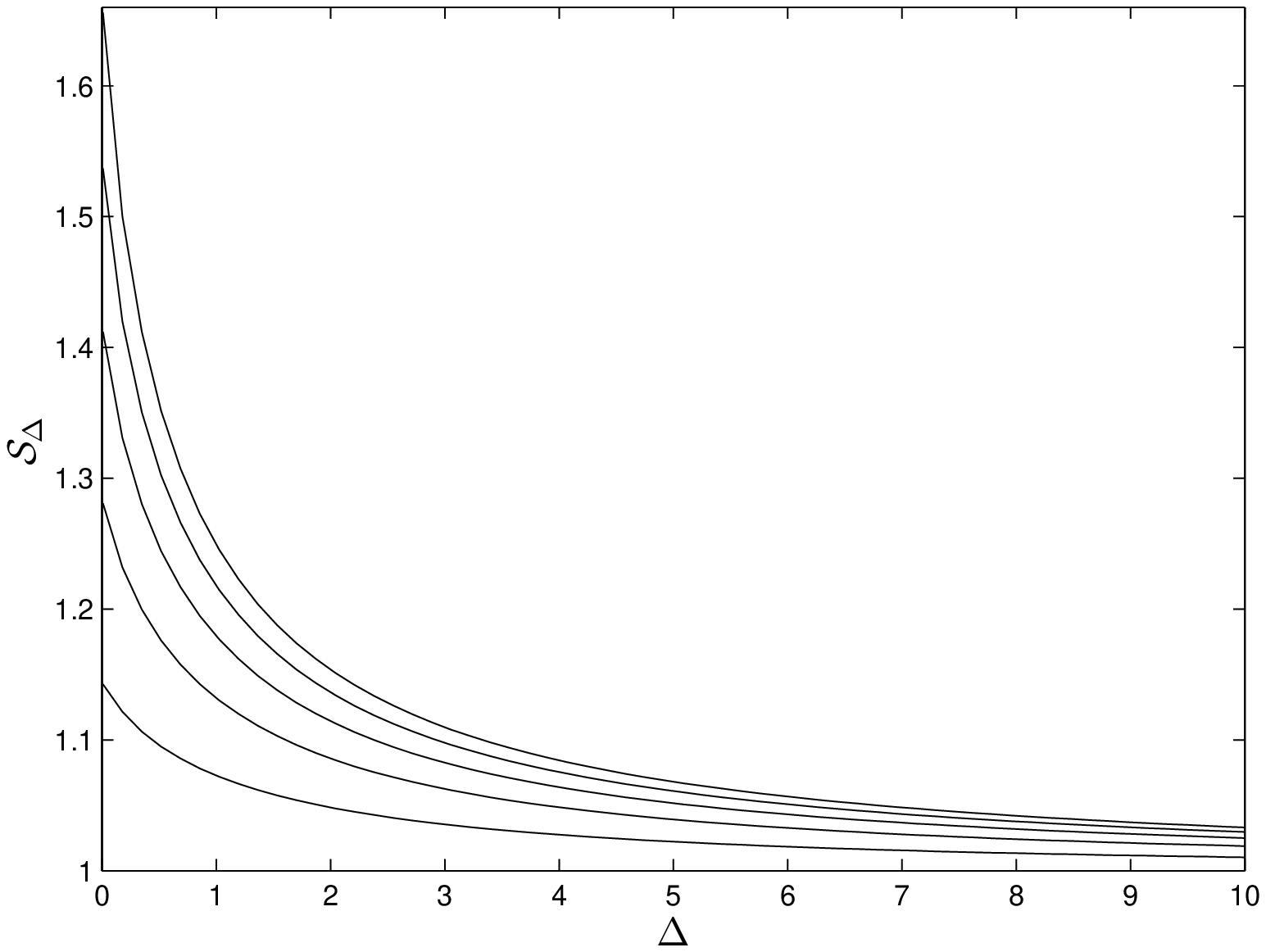}}
{\bf Fig.~12:} \small{Dependence of the generalized strength
$\mathcal{S}_\Delta$ defined by (\ref{fgjujueg}) of the microstructure noise effect
as a function of the interval duration $\Delta$ over which the returns
$R_\Delta(t)$ are defined, for $\alpha=0.1$ ($\sigma=0.2$), $\mu=5$ and
for $q=0; 0.1; 0.2; 0.3; 0.4$ (top to bottom).}
\end{quote}

\clearpage

\begin{quote}
\centerline{
\includegraphics[width=13cm]{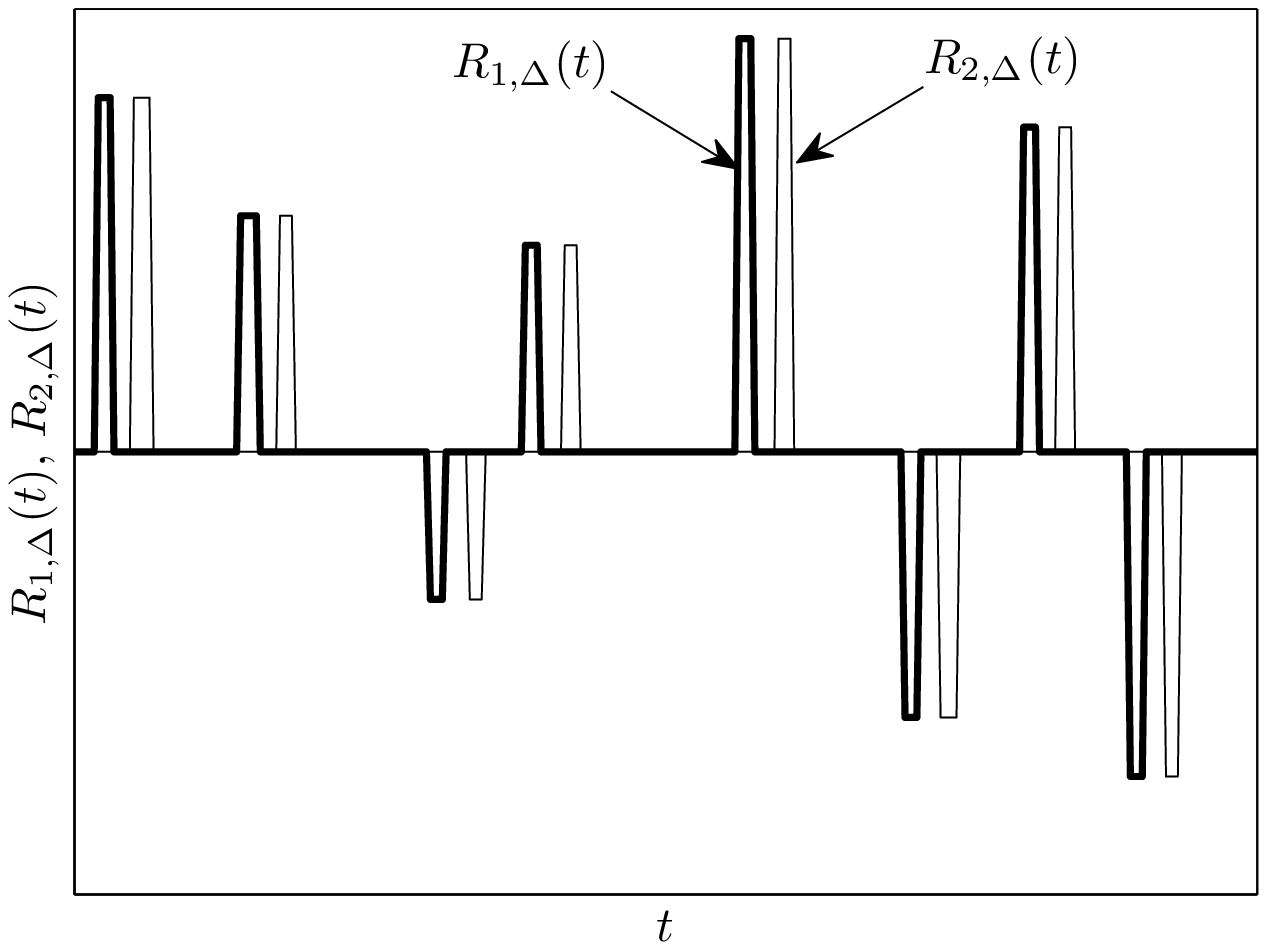}}
{\bf Fig.~13:} \small{A realization of the $\Delta$-scale returns $R_{1,\Delta}(t)$ and $R_{2,\Delta}(t)$ (bold and thin lines correspondingly) for  $\Delta<\zeta$, so that identity \eqref{r12zeroident} is true.}
\end{quote}

\clearpage

\begin{quote}
\centerline{
\includegraphics[width=13cm]{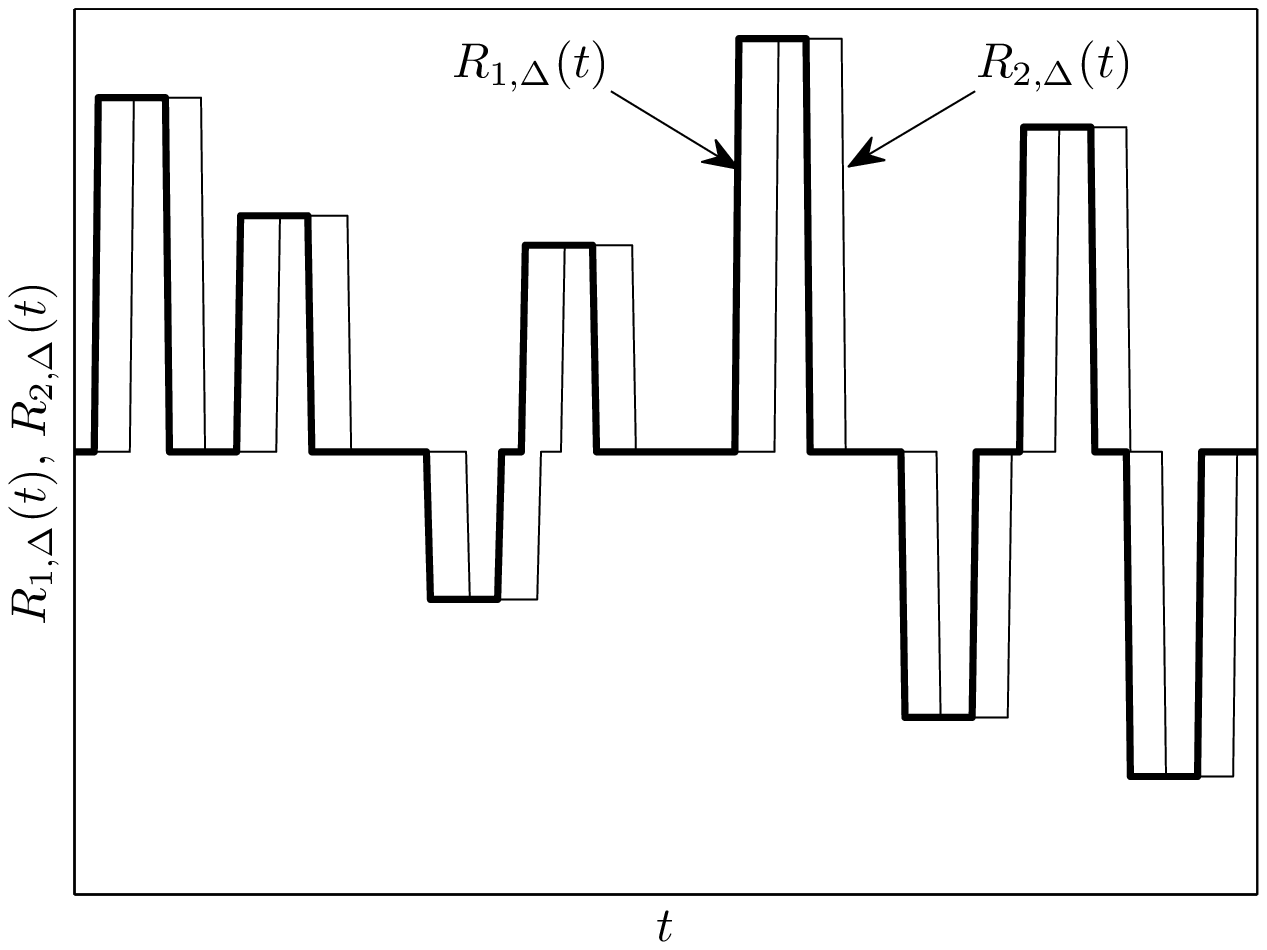}}
{\bf Fig.~14:} \small{A realization of the $\Delta$-scale returns
$R_{1,\Delta}(t)$ and $R_{2,\Delta}(t)$ for $\Delta>\zeta$, so that equality \eqref{dellargepsepps} holds.}
\end{quote}

\clearpage

\begin{quote}
\centerline{
\includegraphics[width=13cm]{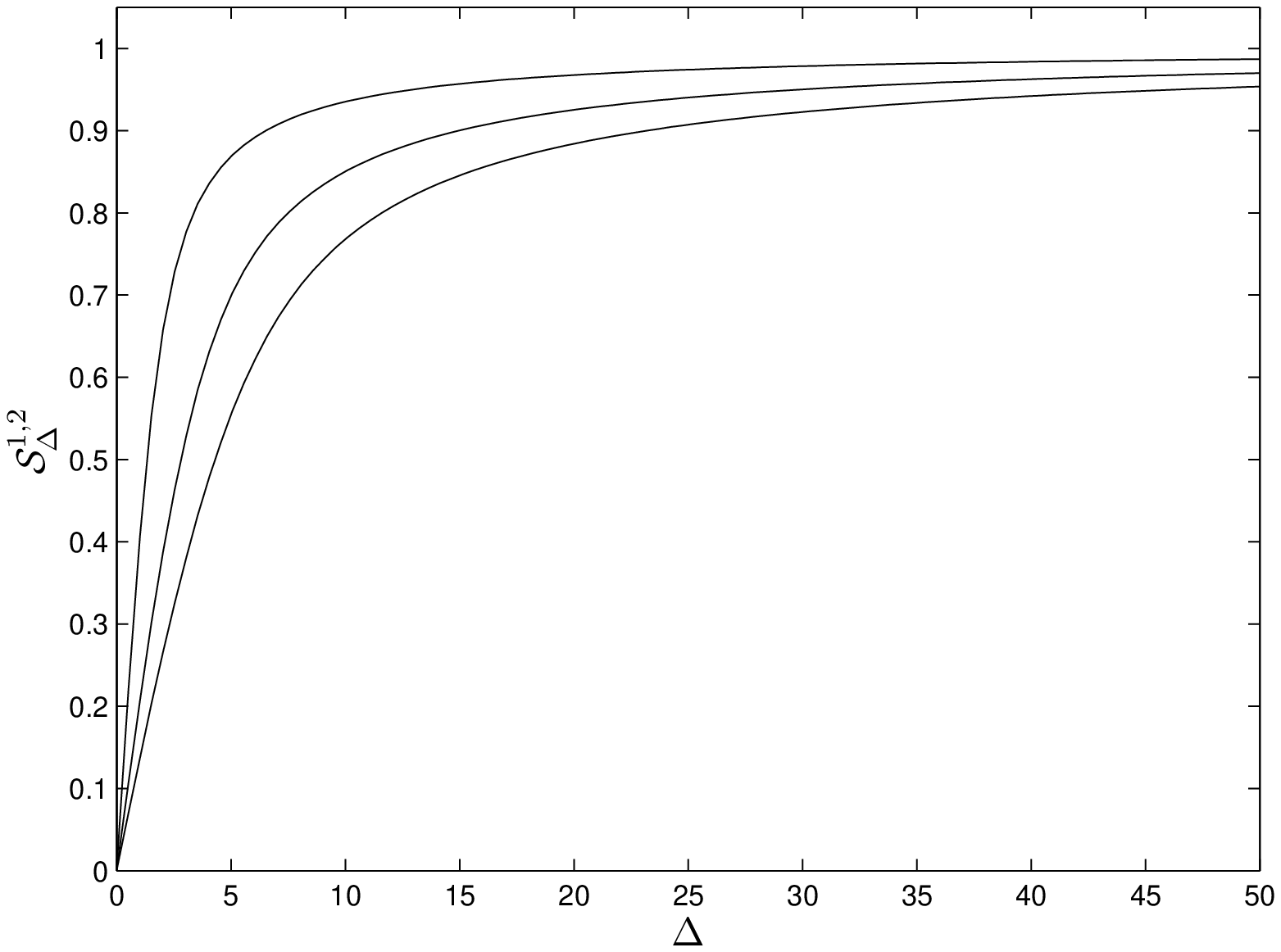}}
{\bf Fig.~15:} \small{Demonstration of the Epps effect in our microstructure model with the dependence of the function $\mathcal{S}^{1,2}_\Delta$ defined in \eqref{rhocross} as a function of $\Delta$, for $\alpha=0.1$ ($\sigma=0.2$), $\mu=4$ and $q=0$.
Top to bottom:  $\lambda = 1; 2; 3$.}
\end{quote}

\clearpage

\begin{quote}
\centerline{
\includegraphics[width=13cm]{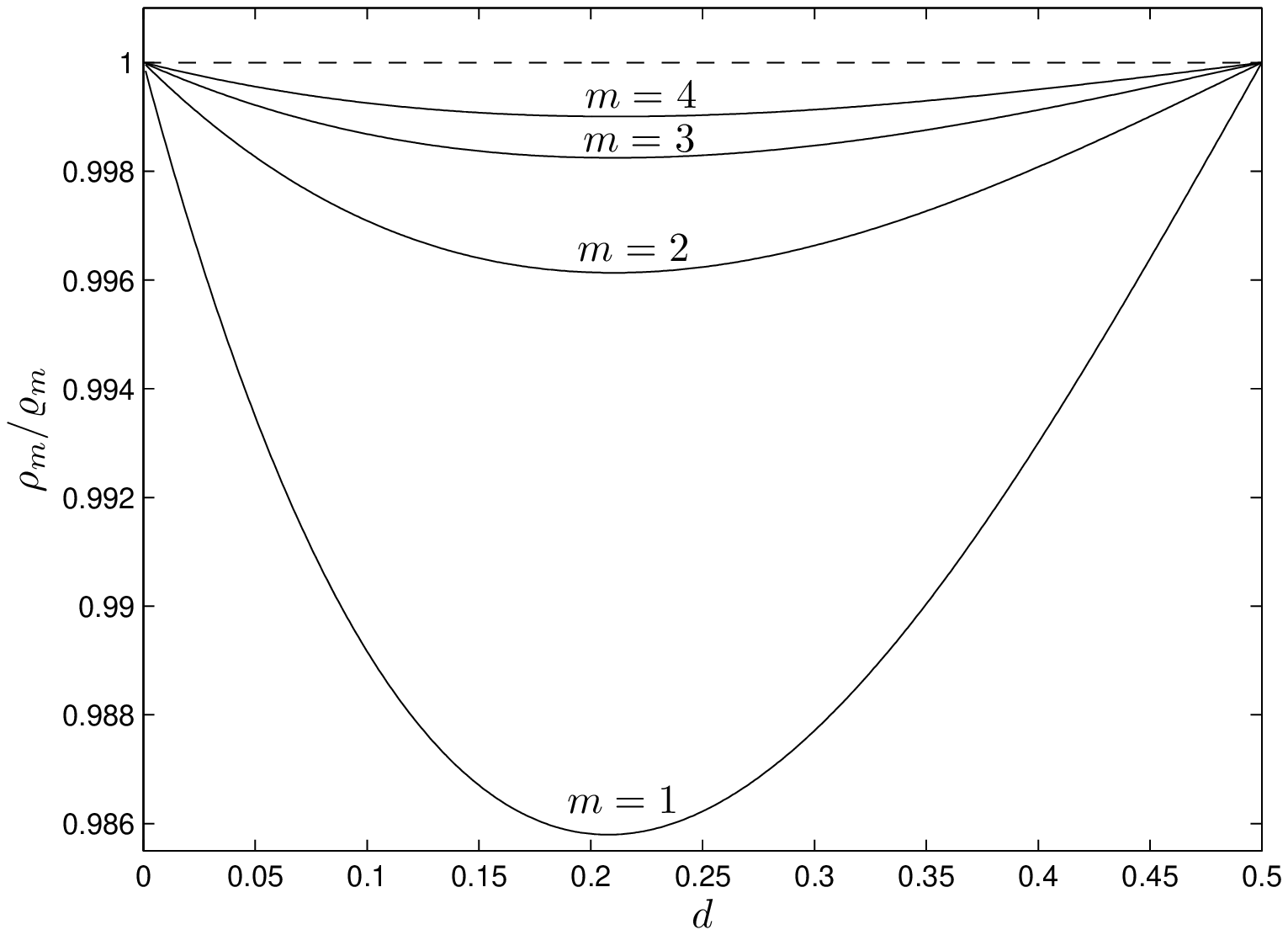}}
{\bf Fig.~A1:} \small{$d$-dependence of the ratio $\rho_m\large/\varrho_m$ of the correlation function $\rho_m$ and its power law approximation $\varrho_m$ for $d\in(0,0.5)$, illustrating the uniform accuracy of the approximation \eqref{rhompower}.}
\end{quote}

\clearpage

\begin{quote}
\centerline{
\includegraphics[width=13cm]{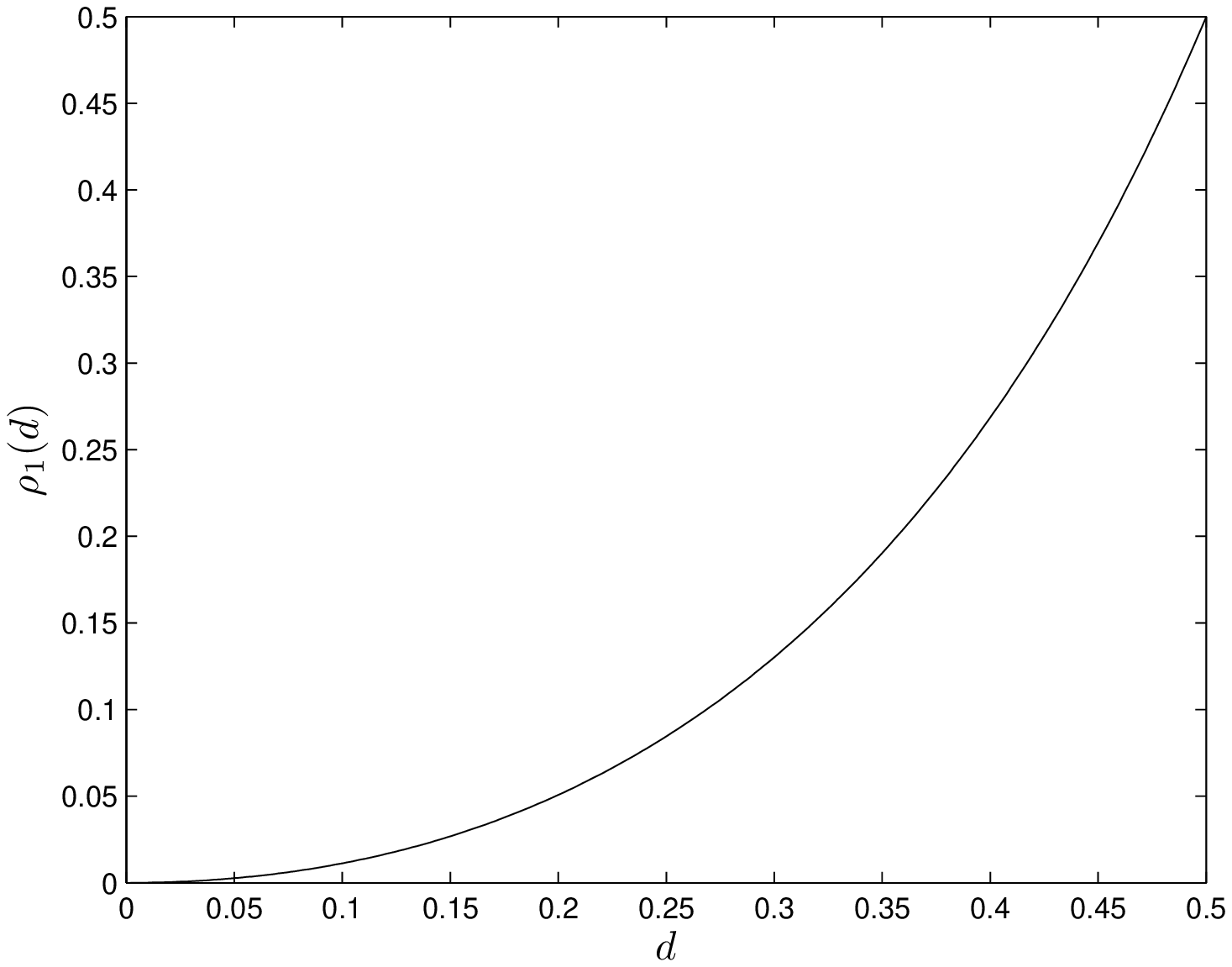}}
{\bf Fig.~A2:} \small{Dependence of $\rho_1(d)$ defined
in (\ref{ehju6pjumj}) as a function of $d$.}
\end{quote}

\clearpage

\begin{quote}
\centerline{
\includegraphics[width=13cm]{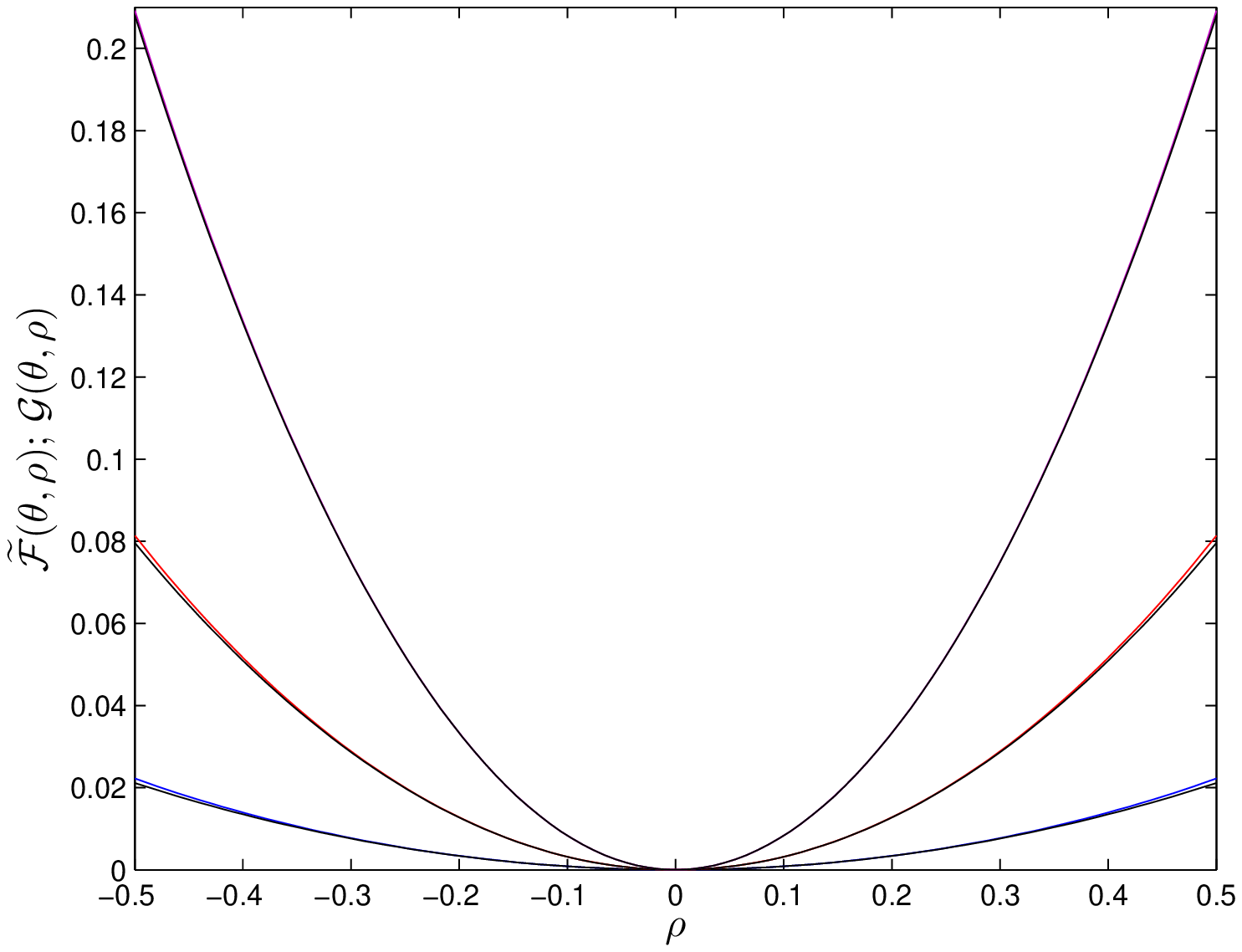}}
{\bf Fig.~A3:} \small{Comparisons of the function $\widetilde{\mathcal{F}}(\theta,\rho)$ \eqref{ftildef}
with its quadratic approximation $\mathcal{G}(\theta,\rho)$ \eqref{gthetrhosq}
as a function of $\rho$. Bottom to top:  $\theta=0.5; 1; 1.5$. For each $\theta$, the
two functions are almost indistinguishable.}
\end{quote}

\clearpage

\begin{quote}
\centerline{
\includegraphics[width=13cm]{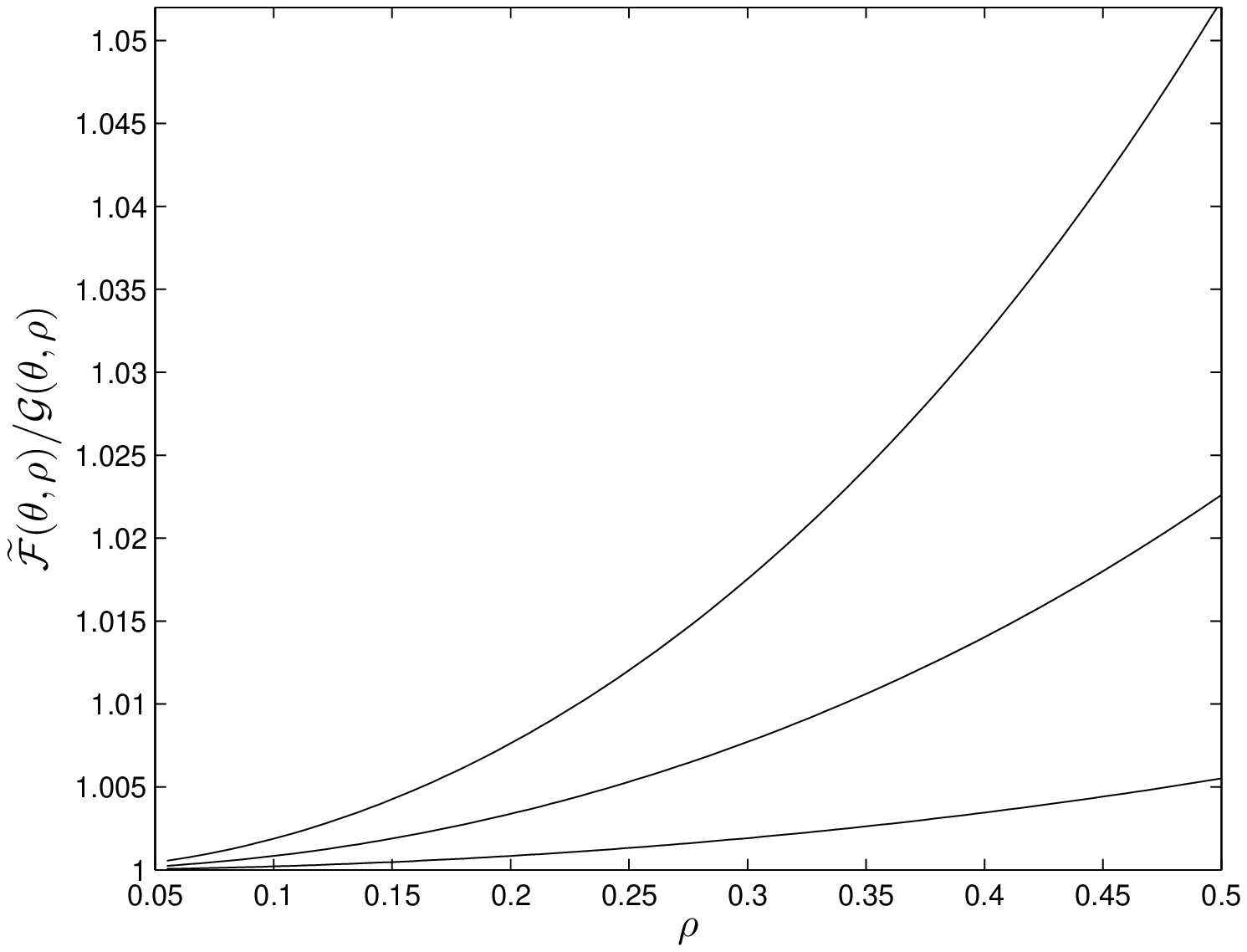}}
{\bf Fig.~A4:} \small{Ratio of the function $\widetilde{\mathcal{F}}(\theta,\rho)$ and its quadratic approximation $\mathcal{G}(\theta,\rho)$ as a function of $\rho$, for all possible $\rho_m$ ($m>0$)  values
as given by condition \eqref{rhomineq}, illustrating the accuracy of quadratic approximation \eqref{gthetrhosq}. Top to bottom: $\theta=0.5; 1; 1.5$.}
\end{quote}


\clearpage

\begin{thebibliography}{99}

\bibitem{Ait2009} A\"it-Sahalia, Y., Yu, J., 2009, High Frequency Market Microstructure Noise Estimates and Liquidity Measures. \emph{The Annals of Applied Statistics}, \textbf{3}, 422--457

\bibitem{Andersen1999} Andersen, T. G., Bollerslev, T., Diebold, F. X., Labys P., 1999, Realized Volatility and Correlation.

\bibitem{Bacry2010} Bacry, E., Delattre, S., Hoffmann, M.,  Muzy, J.-F., 2010,  Modelling microstructure noise
with mutually exciting point processes. \emph{Submitted to Quantitative Finance}

\bibitem{Bandi2006} Bandi, F. M., Russel, J. R., 2006, Separating microstructure noise from volatility. \emph{Journal of Financial Economics}, \textbf{79}, 655--692

\bibitem{Bandi2008} Bandi, F. M., Russel J. R., 2008, Microstructure Noise, Realized
Variance, and Optimal Sampling. \emph{Review of Economic Studies}, \textbf{75}, 339--369

\bibitem{Beran} Beran, J., Statistics for Long-Memory Processes,
Chapman \& Hall/CRC Monographs on Statistics \& Applied Probability (1994)

%\bibitem{Bouchaud2001} Bouchaud, J-P., Matacz, A., Potters, M., 2001, Leverage Effect in Financial %Markets: The Retarded Volatility Model. \emph{Physical Review Letters}, %\textbf{87}, 228701-1-4

\bibitem{Bouchaud2006} Bouchaud, J-P., Potters, M.,  2006, \emph{Financial Risk and Derivative Pricing} (Cambrdidge, UK: Cambridge University Press)

\bibitem{Cont2000} Cont, R., 2001, Empirical properties of asset returns: stylized facts and statistical issues. \emph{Quantitative Finance}, \textbf{1}, 223--236

\bibitem{Cont2010} Cont, R., Stoikov, S., Talreja, R., 2010, A Stochastic Model for Order Book Dynamics. \emph{Operations Research}, \textbf{58}, 549--563

\bibitem{Ding1993} Ding, Z., Grander, C. W. J., Engle, R. F., 1993, A long memory property of stock market returns and a new model. \emph{Journal of Empirical Finance}, \textbf{1}, 83--106

\bibitem{Epps1979} Epps, T. W., 1979, Comovements in Stock Prices in the Very Short Run. \emph{Journal of the American Statistical Association}, \textbf{74}, 291--298.

\bibitem{Munnixetal2011} M\"unnix, M.C., R. Sch\"afer and T. Guhr, 2011.
Statistical causes for the Epps effect in microstructure noise, 
International Journal of Theoretical and Applied Finance 14 (8), 1231-1256

\bibitem{Granger1980} Granger, C. W. J., Joyeux, R., 1980, An Introduction to Long-Memory Time Series Models and Fractional Differencing. \emph{Journal of Time Series Analysis}, \textbf{1}, 15--29.

\bibitem{Hosking1981} Hosking J. R. M., 1981, Fractional Differencing. \emph{Biometrica}, \textbf{68}, 165--176

\bibitem{Ivanov2004} Ivanov, P. Ch., Yuen, A., Podobnik, B., Lee, Y., 2004, Common scaling patterns in intertrade times of U. S. stocks. \emph{Phys. Rev. E}, \textbf{69}, 056107--1--7

%\bibitem{Mainardi2000}

%Mainardi, F., Raberto, M., Gorenflo R., Scalas, E., 2000, Fractional calculus %and continuous-time Finance II:
%the waiting-time distribution. \emph{Physica A}, \textbf{287}, 468--481

%\bibitem{Mike2004}

%Mike, S., Farmer, J. D., 2007, An empirical behavioral model of liquidity and
%volatility. \emph{Journal of Economic Dynamics \& Control}, \textbf{32}, %200--234

\bibitem{Munnix2010} M\"unnix, M. C., Sch\"afer, R., Guhr, T., 2010, Impact of the tick-size on financial returns and correlations. \emph{Physica A}, \textbf{389}, 4828--4843

\bibitem{Politi2008} Politi, M., Scalas, E., 2008, Fitting the Empirical Distribution of Intertrade Durations. \emph{Physica A}, \textbf{387}, 2025--2034

\bibitem{Rhee1997} Rhee S. C., Wang, C.--J., 1997, The Bid-Ask Bounce Effect and the Spread Size Effect: Evidence from the Taiwan Stock Market. \emph{Pacific-Basin Finance Journal}, \textbf{5}, 231--258

\bibitem{Sazuka2007} Sazuka, N., 2007, On the gap between an empirical distribution and an exponential distribution of waiting times for price changes in a financial market. \emph{Physica A}, \emph{376}, 500--506

%\bibitem{Sazuka2009}

%Sazuka, N., Inoue, J., Scalas, E., 2009, The distribution of first-passage %times and durations in FOREX
%and future markets. \emph{Physica. A}, \textbf{388}, 2839--2853

\bibitem{Scalas2007} Scalas, E., 2007, Mixtures of compound Poisson processes as models
of tick-by-tick financial data. \emph{Chaos Solutions \& Fractals}, \textbf{34}, 33--40


\bibitem{Toth2009} T\'oth, B., Kert\'esz, J., 2009, The Epps Effect Revisited. \emph{Quantitative Finance}, \textbf{9}, 793--802

\bibitem{Voev2007} Voev, V., Lunde, A., 2007, Integrated Covariance Estimation using
High-Frequency Data in the Presence of
Noise. \emph{Journal of Financial Econometrics}, \textbf{5}, 68--104

\bibitem{Zhang2005} Zhang, L., Mykland, P.A., and A\"it-Sahalia, Y., 2005, A Tale of Two Time Scales:
Determining Integrated Volatility with Noisy High-Frequency Data. \emph{Journal
of the American Statistical Association}, 100, 1394-1411


\end{thebibliography}
\end{document}